\documentclass[traditabstract]{aa}
\usepackage{graphicx}
\usepackage{natbib}
\usepackage{amssymb}
\usepackage{amsfonts}
\usepackage{amsbsy}
\usepackage{amsmath}

\bibpunct{(}{)}{;}{a}{}{,}

\newcommand{\PFA}{P_{\rm FA}}
\newcommand{\PD}{P_{\rm D}}

\newcommand{\zerob}{\boldsymbol{0}}
\newcommand{\oneb}{\boldsymbol{1}}

\newcommand{\ab}{\boldsymbol{a}}
\newcommand{\bb}{\boldsymbol{b}}
\newcommand{\Bb}{\boldsymbol{B}}
\newcommand{\cb}{\boldsymbol{c}}
\newcommand{\eb}{\boldsymbol{e}}
\newcommand{\fb}{\boldsymbol{f}}

\newcommand{\gb}{\boldsymbol{g}}
\newcommand{\nb}{\boldsymbol{n}}
\newcommand{\pb}{\boldsymbol{p}}
\newcommand{\qb}{\boldsymbol{q}}
\newcommand{\ssb}{\boldsymbol{s}}

\newcommand{\xb}{\boldsymbol{x}}
\newcommand{\wb}{\boldsymbol{w}}
\newcommand{\zb}{\boldsymbol{z}}
\newcommand{\Ab}{\boldsymbol{A}}
\newcommand{\Cb}{\boldsymbol{C}}
\newcommand{\Hb}{\boldsymbol{H}}
\newcommand{\Ib}{\boldsymbol{I}}
\newcommand{\Lb}{\boldsymbol{L}}
\newcommand{\Mb}{\boldsymbol{M}}
\newcommand{\Pb}{\boldsymbol{P}}

\newcommand{\Xb}{\boldsymbol{X}}

\newcommand{\lambdab}{\boldsymbol{\lambda}}
\newcommand{\alphab}{\boldsymbol{\alpha}}
\newcommand{\deltab}{\boldsymbol{\delta}}
\newcommand{\sigmab}{\boldsymbol{\sigma}}
\newcommand{\Deltab}{\boldsymbol{\Delta}}

\newcommand{\ahb}{\boldsymbol{\widehat{a}}}
\newcommand{\phb}{\boldsymbol{\widehat{p}}}
\newcommand{\dhb}{\boldsymbol{\widehat{d}}}

\newcommand{\Smatb}{\boldsymbol{{\mathcal S}}}
\newcommand{\Xmatb}{\boldsymbol{{\mathcal X}}}

\newcommand{\Cmatb}{\boldsymbol{{\mathcal C}}}

\newcommand{\Fmatb}{\boldsymbol{{\mathcal F}}}
\newcommand{\Gmatb}{\boldsymbol{{\mathcal G}}}
\newcommand{\Nmatb}{\boldsymbol{{\mathcal N}}}
\newcommand{\Pmatb}{\boldsymbol{{\mathcal P}}}

\newcommand{\Zmatb}{\boldsymbol{{\mathcal Z}}}

\begin{document}

   \title{An approach for the detection of point sources \\ in very high-resolution microwave maps}

   \author{ Roberto Vio\inst{1}
   \and  Paola Andreani \inst{2} 
 	Elsa Patr\'icia R. G. Ramos\inst{2,3,4} \and Antonio da Silva\inst{3}}
   \institute{ Chip Computers Consulting s.r.l., Viale Don L.~Sturzo 82,
              S.Liberale di Marcon, 30020 Venice, Italy\\
              \email{robertovio@tin.it},
        \and
 	     ESO, Karl Schwarzschild strasse 2, 85748 Garching, Germany\\              
		\email{pandrean@eso.org}
        \and
	     Centro de Astrof\'isica, Universidade do Porto, Rua das Estrelas, 4150-762 Porto, Portugal\\
             \email{eramos@astro.up.pt}      \\
 	     \email{asilva@astro.up.pt}       
        \and
	     Departamento de F\'isica e Astronomia, Faculdade de Ci\^encias, Universidade do Porto, Rua do Campo Alegre, 4169-007 Porto, Portugal
             }

\date{Received .............; accepted ................}

\abstract{This paper deals with the detection problem of extragalactic point sources in multi-frequency, microwave sky maps that will be obtainable in future
cosmic microwave background radiation (CMB) experiments with  instruments capable of very high spatial resolution.
With spatial resolutions  that can be $0.1$-$1.0~{\rm arcsec}$ or better, the extragalactic point sources will appear isolated.
The same also holds for  the compact structures due to the Sunyaev-Zeldovich (SZ) effect (both thermal and kinetic). This situation is different from 
the maps obtainable with instruments such as WMAP or PLANCK where, because of the lower spatial resolution ($\approx 5$-$30 ~{\rm arcmin}$),
the point sources and the compact structures due to the SZ  effect form a uniform noisy background (``{\it confusion noise}''). The point source detection techniques developed in 
the past are therefore based on the assumption that all the emissions
that contribute to the microwave background can be modeled with homogeneous and isotropic (often Gaussian) random fields and make use of the corresponding spatial power spectra.
In the case of very high-resolution observations, such an assumption cannot be adopted since it still holds only for the CMB.
Here, we propose an approach based on the assumption that the diffuse emissions that contribute to the microwave background can be locally approximated by two-dimensional low-order polynomials. In particular, two sets of numerical techniques are
presented that contain two different algorithms each.
The first set makes use of the {\it a-priori} information about the spectral properties of CMB and SZ and is suited to detecting an extragalactic point source with a different spectrum for these emissions. In this set, 
one algorithm is a modification of the {{\it internal linear combination}} (ILC) method, which is widely used in cosmology to extract the component of interest from a mixture of signals, and it is appropriate for extragalactic point sources with a known spectrum. 
The other one does not make use of this piece of information. The second set is tailored to detecting of extragalactic point sources with a similar spectrum to that of the CMB or SZ. Also in this set one algorithm
is specific for extragalactic point sources with known spectrum whereas the other does not make use of this information.
The performance of the algorithms is tested with numerical experiments that mimic the physical scenario expected for high Galactic latitude observations with  the Atacama Large Millimeter/Submillimeter Array (ALMA).}
\titlerunning{point source Detection}
\authorrunning{R. Vio, \& P. Andreani}
\maketitle

\section{Introduction}

The detection of extragalactic point sources in experimental microwave maps is a critical step in analyzing of the {\it cosmic microwave background} (CMB) maps.
Besides the specific interest related to
constructing of dedicated catalogs, these sources can, if not properly removed, have adverse effects on the estimation of the power spectrum and/or the test of Gaussianity of the CMB
component. Much effort has been dedicated to multiple frequency maps of the same sky area, and
many algorithms have been proposed \citep[see][ and references therein]{her08, her12}. Apart from a recent Bayesian approach \citep{car09}, most of them belong to two broad classes of techniques. 
The first class, suited to the extragalactic point sources with known spectra,  is based on the  {\it Neyman-Pearson} (NP)
criterion that consists in maximizing the {\it probability of detection} $\PD$ under the constraint that the {\it probability of false alarm} $\PFA$ (i.e., the probability of a false detection) does not exceed a fixed value
$\alpha$ \citep{kay98}. The resulting algorithms are extensions of the classic matched filter (MF)  \citep{her02, ram11}. The second class, appropriated to the extragalactic point sources with unknown spectra, is based on maximizing
of the ``{\it signal-to-noise ratio}'' (${\rm S/N}$) of the source intensity with respect to the underlying background \citep{ her09, lan10}. Both classes require the spatial power spectrum of the emitting components 
and therefore are based on the assumption that the emissions contributing to the
microwave sky can be modeled by means of homogeneous and isotropic random fields. 

With a spatial resolution worse than $5~{\rm arcmin}$ typical of Planck and WMAP experiments, such an assumption is excellent for the CMB, 
good for the {\it confusion noise} due to the point sources and the Sunyaev-Zeldovich (SZ) effect component, and locally barely acceptable for the diffuse Galactic emission. In the case of observation at very high spatial resolution, 
of order of $0.1$-$1.0 ~{\rm arcsec}$ (that will be possible, for instance, with instruments as ALMA), this is no longer true. Indeed,
since almost all of the extragalactic point sources but also of the compact structures due to the SZ effect will appear isolated, they can be thought of as the realization of (not necessarily stationary) shot-noise processes. Moreover,
on small observing areas, it is realistic to expect that  any other additional diffuse component (e.g. due to the Galactic emissions) is almost constant or slowly changing. This experimental scenario is different from the
past or ongoing CMB experiments. New techniques of point source detection are therefore necessary. 

From the consideration that the detection of extragalactic point sources is typically done on very small areas of the sky, where the contribution of the diffuse components
can be approximated well by means of low-degree, two-dimensional polynomials, we propose an approach based on two sets of algorithms. The first set makes use of the {\it a-priori} information about the spectral properties of 
CMB and SZ, and the second one does not exploit this piece of information.
Each set contains two algorithms that are suited to detecting point sources with known and unknown spectra, respectively. The reason for two different sets of algorithms is that the use of the spectral properties of the CMB and SZ permits
removal of the contribution from these components. As a consequence, the knowledge of their spatial power spectrum is no longer necessary, and at the same time it is possible to unambiguously distinguish between true extragalactic 
point sources and the compact structures due to SZ. The price is a reduced, or even null,  detection capability for the point sources with spectra that are similar, or even identical, to that of CMB, SZ, or to a linear combination of these. 
The algorithms that do no make use of this piece of information do not suffer this limitation, but they cannot distinguish the true extragalactic point sources from the compact structures due to SZ. 

The reason for two algorithms within each set is that those specialized for the extragalactic point sources with a specific spectrum are characterized by a greater detection capability.  They 
are obtained by modifying the {\it internal linear combination} (ILC) method that in cosmology is used to extract a  component of interest from the mixture of signals that contribute to the microwave sky emission \citep{eri04, hin07, vio08}.
Their main limitation is the necessity of multiple applications for detecting point sources with different spectra. Such a necessity is avoided by the algorithms that do not exploit this piece of information but at the price of
less detection capability. The combined use of a couple of these algorithms belonging to different sets permits the detection of extragalactic point sources independently of their spectral emission and reduced contamination due to the
SZ compact structures. The performances of both sets of algorithms is tested via numerical experiments based on simulated maps of high Galactic latitude that might be the area of interest of CMB high spatial resolution observations.

The paper is organized as follows. Section 2 introduces the mathematical framework and explains the algorithms used. Section 3 discusses practical problems
related to the choice of the experimental parameters. Numerical experiments are reported in section 4, while conclusions are summarized in section 5.

\section{Formalization and  solution of the problem} \label{sec:detection}

By searching for a single extragalactic point source in a small area of sky, the microwave emission can be modeled with bidimensional discrete patches $\{ \Xmatb_i \}_{i=1}^{N_f}$, each of them containing
$N_p = N_{p_1} \times N_{p_2}$ pixels, corresponding to $N_f$ different observing frequencies (channels), with the form
\begin{equation} \label{eq:observed2}
\Xmatb_i = \Smatb_i + \Cmatb_i +  \Zmatb_i + \Gmatb_i + \Nmatb_i.
\end{equation} 
Here, $\Smatb_i$ is the contribution of the extragalactic point source at the $i$th frequency, $\Cmatb_i$,  $\Zmatb_i$, and  $\Gmatb_i$ are the backgrounds due to CMB, extragalactic emission, 
and some other possible diffuse component (e.g. Galactic emission), respectively,
and  $\Nmatb_i$ is the instrumental noise. In this model, the contribution of the extragalactic point sources is assumed in the form
\begin{equation} \label{eq:point}
\Smatb_i = a_i \Fmatb,
\end{equation}
with ``$a_i$'' the intensity of the source at the $i$th channel. According to 
Eq.~(\ref{eq:point}), and without loss of generality, all the sources are assumed to have the same profile $\Fmatb$ independently of the 
observing frequency. In practical applications, this is not true. However, it is possible to meet this condition by convolving the images with an appropriate kernel (see below). 

For computational reasons that soon will become evident, it is useful to convert the two-dimensional model~(\ref{eq:observed2}) into the one-dimensional form
\begin{equation} \label{eq:observed1}
\xb_i = \ssb_i + \cb_i + \zb_i + \gb_i + \nb_i.
\end{equation}
Here, $\xb_i =  {\rm VEC}[\Xmatb_i]$, with  ${\rm VEC}[ \Hb]$ the operator that transforms a matrix $\Hb$ into a vector by stacking its columns one underneath the other. Something similar holds for the other quantities.

\subsection{Detection with ILC background removal} \label{sec:removal1l}

One classical solution to deal with many maps of the same sky area taken at different observing frequencies consists in a linear composition
by means a set of weights $\wb=[w_1, w_2, \ldots, w_{N_f}]^T$. In this way it is possible to work with a single map given by
\begin{equation}
\xb = \Xb \wb,
\end{equation}
where $\Xb = ( \xb_1, \xb_2, \ldots, \xb_{N_f})$ is an $N_p \times N_f$ matrix.
The obvious question is how to fix such weights.

Before proceeding, it is necessary to take into account that there is an {\it a-priori} information about the various components in Eq.~(\ref{eq:observed2}). 
In particular,
\begin{itemize} 
\item For each observing frequency $i$, the spectra of $\Cmatb_i$ and $\Zmatb_i$ are known with good accuracy. Moreover, the spatial distribution of these components is independent of the observing frequency;\\
\item $\Cmatb_i$ has a diffuse spatial distribution with a coherence scale of about $10~{\rm arcmin}$, which is much greater than that of the point sources. It is reasonable to expect that something similar could hold for $\Gmatb_i$; \\
\item Noises $\Nmatb_i$ can be reasonably assumed as given by the realization of independent Gaussian white-noise processes with standard deviation $\sigma_{\Nmatb_i}$.
\end{itemize}
The first point implies that, for a given observing frequency, the emission due to the CMB and the SZ components can be obtained from a rescaled linear mixture of two templates (i.e., maps that do not depend on frequency) 
$\cb$ and $\zb$, respectively. This implies that the cumulative contribution $\bb_i = \cb_i + \zb_i$ of these components is given by the $i$th column of matrix
\begin{equation}
\Bb = (\cb, \zb) \Mb,
\end{equation}
where $\Mb$ is an $N_e \times N_f$ matrix usually indicated with the term of {\it mixing matrix}. In the present context, the number of emission mechanisms is $N_e = 2$ since the kinetic SZ emission has the same spectrum as for the CMB. For this reason,  with $\cb_i$ we indicate the CMB plus the kinetic SZ emission from now on. The second point implies that within a small area centered on an extragalactic point source, the CMB and any other diffuse emission
vary very little. This suggests that, for any patch $\Xmatb_i(j,k)$ with $-N_j \le j \le N_j$ and $-N_k \le k \le N_k$ ($N_{p_1} = 2 N_j+1$, $N_{p_2}= 2 N_k+1$),
these emissions can be safely approximated by a low-degree, two-dimensional polynomial of degree $m$
\begin{equation}
\Pmatb_m (j,k) = \sum_{l=0}^m \alpha_l (j^q k^r); \qquad q + r \le l,
\end{equation}
where $\{ \alpha_l \}$ are real coefficients, whereas $q$ and $r$ are integer numbers permuted accordingly.

\subsubsection{Point sources with known spectrum} \label{sec:mmilc}

Starting from these considerations and adopting the criterion of the {\rm S/N} maximization  for a given extragalactic point source with emission spectrum
 $\ab = a \ahb$, where ``$a$'' is to be estimated and $\ahb= [\widehat{a}_1, \widehat{a}_2, \ldots, \widehat{a}_{N_f}]^T$ is fixed, the weights $\wb$ can be computed 
through the maximization of the quantity  
\begin{equation} \label{eq:model1}
{\rm SNR} = \frac{(\ahb^T \wb)^2}{ \Vert ( \Xb \wb - \Lb \qb) \Vert^2};
\end{equation}
with ``$\Vert . \Vert$'' the Euclidean norm, under the constraints
\begin{align}
\ahb^T \wb &=1;  \label{eq:constr1} \\
\Mb \wb &= \label{eq:constr2}
\left( \begin{array}{c}
0 \\
0
\end{array} \right).
\end{align}
Because of  the constraint~(\ref{eq:constr1}), which forces weights $\wb$ to preserve the intensity $a$, the numerator of Eq.~(\ref{eq:model1}) is a constant. Therefore, the maximization of {\rm SNR} is obtained by the minimization of the denominator.
The rationale behind this approach is that if each of the maps $\xb_i$ contains the contribution of a point source with shape $\Fmatb$ and of 
smooth component $\gb_i$, approximable by means of a two-dimensional  polynomial, then the same has
to hold for their linear combination $\xb=\Xb \wb$. The denominator $\xb - \Lb \qb$ thus represents the residuals of the least-squares fit to the composite map $\xb$ of a model where a central point spread function (PSF) is superimposed
to a bivariate polynomial background. The weights $\wb$ are computed in such a way as to minimize the standard deviation of these residuals under the constraints~(\ref{eq:constr1}) and (\ref{eq:constr2}).   
In Eq.~(\ref{eq:model1}), $\qb=(a, \alphab^T)^T$ is an array with size $N_c=[(m+1) (m+2)/2] +1$, $\alphab$ the coefficients of the two-dimensional polynomial,
whereas $\Lb$ is an $N_p \times N_c$ matrix with the form $\Lb = [\fb, \Pb]$ with $\fb = {\rm VEC}[\Fmatb]$ and $\Pb$ the $N_p \times (N_c-1)$ {\it design matrix} corresponding to 
the least-squares fit of a two-dimensional polynomial \footnote{If the degree is one,  $\Pb = [\deltab_1, \deltab_2, \oneb]$, whereas for a degree two 
$\Pb = [\deltab_1 \odot \deltab_1, \deltab_2 \odot \deltab_2, \deltab_1 \odot \deltab_2, \deltab_1, \deltab_2, \oneb]$, where ``$\odot$'' represents the {\it element-wise}
matrix multiplication (Hadamard product), $\oneb$ is a vector of ones, and $\deltab_1 = {\rm VEC}[\Deltab_1]$, $\deltab_2 = {\rm VEC}[\Deltab_2]$, where $\Deltab_1$ is a matrix with $2 N_j+1$ identical columns 
$[-N_k, -N_k+1, \ldots, 0, \ldots, N_k-1, N_k]^T$, whereas $\Deltab_2$ is a matrix with $2 N_k+1$ identical rows $[-N_j, -N_j+1, \ldots, 0, \ldots, N_j-1, N_j]$.}.
The constraint~ (\ref{eq:constr2}) forces the contribution of the CMB and SZ components to the final map $\xb$  to zero, whereas the contamination $\gb$ is removed by means of the two-dimensional polynomial.
The quantities $\wb$ and  $\qb$ are unknown and have to be estimated. The maximization of {\rm SNR} with the constraints~(\ref{eq:constr1})-(\ref{eq:constr2})
can be written in the form 
\begin{align} \label{eq:problem1}
& R(\wb, \qb,\lambdab) =  \nonumber \\ 
&\underset{ \wb, \qb, \lambdab}{\arg\min}  \left[ \Vert ( \Xb \wb - \Lb \qb) \Vert^2 + \lambdab^T (\Mb_a^T \wb - \eb_1) \right],
\end{align}
with  $\Mb_a = (\ahb, \Mb^T)$ , $\lambdab$ an $N_e+1$ array of {\it Lagrange multipliers}, and $\eb_1$ an $N_e + 1$  array of zeros except for the first element, which is ``$1$''. 

This method is a modification of the {\it constrained} ILC by \citet{rem11}. We call it  {\it modified multiple} ILC (MMILC). The basic idea is that, if in the center of the selected patch there is a point source, 
then the value of ``$a$'' should exceed a threshold due to  noise. After some algebra, one obtains that the solution of problem~(\ref{eq:problem1}) is given by the system of equations
\begin{equation} \label{eq:solution1}
\left( \begin{array}{ccc}
+2 \Cb_{XX} & -2 \Cb_{XL} & \Mb_a \\
 -2 \Cb_{XL}  & +2 \Cb_{LL} & \zerob \\
\Mb_a^T & \zerob & \zerob
\end{array} \right) 
\left( \begin{array}{c}
\wb \\
\qb \\
\lambdab
\end{array} \right) =
\left( \begin{array}{c}
\zerob \\
\zerob \\
\eb_1
\end{array} \right),
\end{equation}
$\Cb_{XX} =  \Xb^T \Xb$, $\Cb_{XL} = \Xb^T \Lb$ and $\Cb_{LL}=\Lb^T \Lb$, which provides
\begin{align}
\wb & = \Hb^{-1} \Cb_{XX}^{-1} \Mb_a (\Mb_a^T \Cb_{XX}^{-1} \Mb_a)^{-1} \eb_1, \label{eq:sol1} \\
\qb & = \Cb_{LL}^{-1} \Cb_{XL}^T \wb, \label{eq:sol2}
\end{align}
where
\begin{equation}
\Hb= \left[ \Ib - \Cb + \Cb_{XX}^{-1} \Mb_a (\Mb_a^T \Cb_{XX}^{-1} \Mb_a)^{-1} \Mb_a^T \Cb   \right], \label{eq:sol3}
\end{equation}
with $\Ib$ the identity matrix, and
\begin{equation}
\Cb = \Cb_{XX}^{-1} \Cb_{XL} \Cb_{LL}^{-1} \Cb_{XL}^T. \label{eq:sol4}
\end{equation}
One interesting characteristic of solution~(\ref{eq:solution1}) is that it does not require knowing the noise level of each map, a quantity that often can only be roughly estimated.

A useful insight into how MMILC works can be obtained if problem~(\ref{eq:problem1}) is recast in the form \footnote{We thank the referee for this
suggestion.}
\begin{equation} \label{eq:problem3}
R(\wb) =
\underset{ \wb}{\arg\min}  \Vert \Ab \wb \Vert^2     \qquad \text{subject to: }  (\Mb_a^T \wb - \eb_1) 
\end{equation}
where $\Ab = (\Ib -  \Lb (\Lb^T \Lb)^{-1} \Lb^T) \Xb$. Since matrix $\Ab$ is the orthogonal projection of $\Xb$ onto the {\it nullspace} of $\Lb^T$,  MMILC can be seen to sequentially perform a least-squares fit of the PSF 
overlapping a two-dimensional polynomial background on the original data for each frequency, followed by a constrained ILC on the residuals.

\subsubsection{Point sources with unknown spectrum}

The main limitation of  MMILC is that it only works optimally for a specific emission spectrum $\ahb$. This assumption can be relaxed by converting the maximization of ${\rm SNR}$ with the constraints~(\ref{eq:constr1})-(\ref{eq:constr2}) 
in the least-squares minimization of the quantity
\begin{equation} \label{eq:model1a}
S = \Vert \Xb \wb - \Lb \qb \Vert^2,
\end{equation}
with the constraints
\begin{align}
\wb^T \wb &=1;  \label{eq:constr1a} \\
\Mb \wb &= \label{eq:constr2a}
\left( \begin{array}{c}
0 \\
0
\end{array} \right).
\end{align}
The constraint~(\ref{eq:constr1a}) is set to avoid the trivial solution $\wb = \zerob$. In this way, problem~(\ref{eq:problem1}) is converted into
\begin{align} \label{eq:problem1a}
& S(\wb, \qb,\lambdab) =  \nonumber \\ 
&\underset{ \wb, \qb, \lambdab}{\arg\min}  \left[ \Vert ( \Xb \wb - \Lb \qb) \Vert^2 + 2 \lambdab^T \Mb + \gamma (\wb^T \wb -1)  \right].
\end{align}
After some algebra, it is possible to see that $\qb$ is again given by Eq.~(\ref{eq:sol2}) but with $\wb$ the solution of the eigenvalue problem
\begin{equation} \label{eq:sol3a}
\Hb \wb = \gamma \wb,
\end{equation}
where
\begin{equation}
\Hb = (\Ib - \Mb^T \Cb_{MM}^{-1} \Mb) (\Cb_{XX} - \Cb_{XL} \Cb_{LL}^{-1} \Cb_{XL}^T),
\end{equation}
and $\Cb_{MM} = \Mb \Mb^T$. The searched for $\wb$ is given by the eigenvector of $\Hb$ that minimizes quantity $S$. Presently, the only method that we can suggest is to insert each eigenvector in Eq.~(\ref{eq:model1a}) and
to check numerically which of them provides the smallest $S$. This is because we have ascertained that there are situations where the eigenvector corresponding to the lowest eigenvalue of $\Hb$ (a criterion typical of the least-squares problems) 
does not work. Indeed, matrix $\Hb$ is not symmetric, and it cannot be expected to have any particular property. We call this method {\it nonparametric} MMILC (NP-MMILC).

Although not specifically optimized for a particular $\ahb$, the results provided by NP-MMILC depend on the emission spectrum of the point source. Indeed, if like MILC, this method too is interpreted as a sequential least-squares fits followed by a constrained ILC, it can be understood that, in the case of a point source with a large amplitude in maps with a low ${\rm S/N}$
and a small amplitude in maps with a high ${\rm S/N}$, this results in a reduced detection capability.
A simple procedure for avoiding this problem consists of applying NP-MMILC 
not to all maps but only to those for which the best ${\rm S/N}$ for the point source is expected. The choice can be based on $\ahb$. This means to use the {\it a priori} information on the emission spectrum in a different way from MMILC.

\subsubsection{Detection procedure} \label{sec:threshold}

When searching for extragalactic point sources with MMILC or NP-MMILC in a given set of maps, the procedure consists in fixing the size $(2 N_j +1) \times (2 N_k +1)$  of a window that is made to slide, pixel by pixel, across the area of interest. 
At the end of this procedure a single map is obtained containing the estimated values of ``$a$'' for each pixel. Now,  the question is to fix the detection threshold below which a given value of ``$a$'' is supposed
to be only due to noise. In this respect, the direct use of solutions~(\ref{eq:sol1})-(\ref{eq:sol3}) and  (\ref{eq:sol3a}) is difficult. For this reason, two different procedures are suggested,
\begin{enumerate}
\item For MMILC it is set $a = 0$ if $a \le k \sigma_{{\rm L}}$, where $k$ is a constant factor (typically $k=4,5$), $\sigma_{{\rm L}} = \Vert \sigmab^T_{n}  \wb \Vert \sqrt{(\Lb^T \Lb)^{-1}_{1, 1}}$,  $ \sigmab_{n} =  
( \sigma_{n_1},  \sigma_{n_2}, \ldots,  \sigma_{n_{N_f}})^T$, and
$(\Lb^T \Lb)^{-1}_{1, 1}$ is the first entry of matrix  $(\Lb^T \Lb)^{-1}$. This operation corresponds to estimating the standard deviation $\sigma_a$ of ``$a$'' for a fixed
$\wb$. Such an approach has the advantage that matrix $(\Lb^T \Lb)^{-1}$  can be computed only once since it is the same for all the patches. But, it has the disadvantage that the standard deviations of the noises 
$\{ n_i \}$ have to be known in advance. In the case of NP-MMILC, a similar procedure holds, but it is necessary to take the absolute value of $\ab$. Indeed, it is readily verified that, with $\qb$ given by
Eq.~(\ref{eq:sol2}), a change of sign of $\wb$ does not modify the value of $S$; \\ 
\item  It is set  $a=0$ if $a \le k \sigma_{{\rm map}}$, for MMILC, and $|a| \le k \sigma_{{\rm map}}$, for NP-MMILC. Again, $k$ is a constant factor and $\sigma_{{\rm map}}$ is the standard deviation of the entries in the final map. 
This is an unsophisticated approach, but it does have the advantage
of not requiring the standard deviation of the noise in each patch, a quantity usually known only roughly.
\end{enumerate}

Before concluding this section, we underline that the number of rows $N_e=2$ of the {\it mixing matrix} $\Mb$ comes from our interest in exploring the situation in which the extragalactic component $\Zmatb_i$ 
consists of secondary anisotropies of the CMB. In particular, we have only considered the SZ effect (both thermal and kinetic), which is the strongest one in galaxy clusters, groups of galaxies, and in protoclusters \citep{bir99}.
However, if the information is available for one or more additional components, then it is sufficient to update $\Mb$ and the same solutions~(\ref{eq:sol1})-(\ref{eq:sol3}) and (\ref{eq:sol3a}) and hold for $N_e = 3$ or greater. Similarly,
if one decides to remove only one component via ILC, either CMB or SZ, then it is sufficient to eliminate the appropriate row from matrix $\Mb$ and set $N_e=1$ in the solution. In this way, however, the drawback is that the
remaining component has to  be removed by means of the two-dimensional polynomial. This could be a necessary operation in the case of noisy maps (see below).

\subsection{Detection without ILC background removal}

The MMILC and NP-MMILC detection techniques are potentially quite effective, however they suffer from two main drawbacks:
\begin{enumerate}
\item To remove the CMB and SZ components, one or more of the weights in $\wb$ have to be negative. As a consequence, since in the final map $a = \ab^T \wb$ and $\sigma_{{\rm map}} = 
\Vert \sigmab^T_{n}  \wb \Vert$,  ``$a$'' is given by the sum of both positive and negative values whereas, $\sigma_{{\rm map}}$ is given by the sum of positive values alone. In other words, the background subtraction reduces
 the ${\rm S/N}$ with respect to a simple sum of the maps. The situation worsens when the emission of an extragalactic point source has a spectrum similar to that of the CMB or of the SZ since ``$a$'' will tend to zero; \\
\item If $\ahb$ is an array such that $\Mb \ahb = \zerob$, i.e. $\ahb$ belongs to the {\it nullspace} of $\Mb$  (i.e., it is given by the linear combination of the column of $\Mb$), then the
system~(\ref{eq:solution1}) does not have any useful solution since, when $\wb = \ahb / \Vert  \ahb \Vert^2$, both constraints $\ahb^T \wb =1$ and $\Mb \wb = \zerob$ are satisfied, but it happens that the estimate $a$ is such that its expected value is zero.
\end{enumerate} 
For this reason, to detect extragalactic point sources with $\ahb$ belonging to the {\it nullspace} of $\Mb$, the above procedures have to be adapted to work without the ILC removal of the CMB and the SZ components. Again,
two different algorithms are presented.

\subsubsection{Point sources with known spectrum}

The case of point sources with known spectra can be easily obtained from problem~(\ref{eq:problem1})  through the substitutions $\Mb_a = \ahb$ and  $\eb_1=1$:
\begin{equation} \label{eq:problem2}
R(\wb, \qb, \lambda) =  \underset{ \wb, \qb, \lambda}{\arg\min}  \left[ \Vert ( \Xb \wb - \Lb \qb) \Vert^2 + \lambda (\ahb^T \wb - 1) \right] ,
\end{equation}
with solution given by
\begin{equation} \label{eq:solution2}
\left( \begin{array}{ccc}
+2 \Cb_{XX} & -2 \Cb_{XL} & \ahb \\
-2 \Cb_{XL} & +2 \Cb_{LL} & \zerob \\
\ahb^T & \zerob^T & 0
\end{array} \right) 
\left( \begin{array}{c}
\wb \\
\qb \\
\lambda
\end{array} \right) =
\left( \begin{array}{c}
\zerob \\
\zerob \\
1
\end{array} \right).
\end{equation}
The explicit solution for $\wb$ and $\qb$ is given by Eqs.~(\ref{eq:sol1})-(\ref{eq:sol3}) with $\Mb_a = \ahb$. Detection is still carried out as explained in Sec.~\ref{sec:threshold}.
With this method, which we call {\it modified} ILC (MILC), the CMB  and SZ emissions are not removed through the use of the {\it mixing matrix} $\Mb$, but rather by exploiting that the CMB, part of the SZ, and any other component
with a diffuse spatial distribution can be removed through the polynomial
approximation. As a consequence, the only contribution in the final map beyond that of the extragalactic point sources is the compact component of the SZ (both thermal and kinetic), and this is an unavoidable problem. Without
additional information, it is impossible to separate an SZ emission with point-like spatial distribution from a genuine extragalactic point source. In the case of SZ emission with more extended structures, a possible solution consists of checking if their spatial distribution is compatible with the PSF $\Fmatb$. This issue, however, is beyond the scope of the present work.

\subsubsection{Point sources with unknown spectrum}

If in the minimization of the quantity $S$ as given in Eq.~(\ref{eq:model1a}) the constraint~(\ref{eq:constr2a}) is relaxed, the {\it nonparametric} version of MILC is obtained (NP-MILC). It is easily verified that  for this problem too, $\qb$
is given by  Eq.~(\ref{eq:sol2}), but now $\wb$ is the solution of the eigenvalue problem $\Hb \wb = \gamma \wb$ with
\begin{equation}
\Hb =  (\Cb_{XX} - \Cb_{XL} \Cb_{LL}^{-1} \Cb_{XL}^T).
\end{equation}
As for NP-MMILC, the searched $\wb$ is given by the eigenvector of $\Hb$ that minimizes quantity $S$, and detection is carried out as explained in Sec.~\ref{sec:threshold}. Also, NP-MILC suffers the same dependence
on $\ahb$ as NP-MMILC.

\section{Practical uses}
 
\begin{figure*}
        \resizebox{\hsize}{!}{\includegraphics{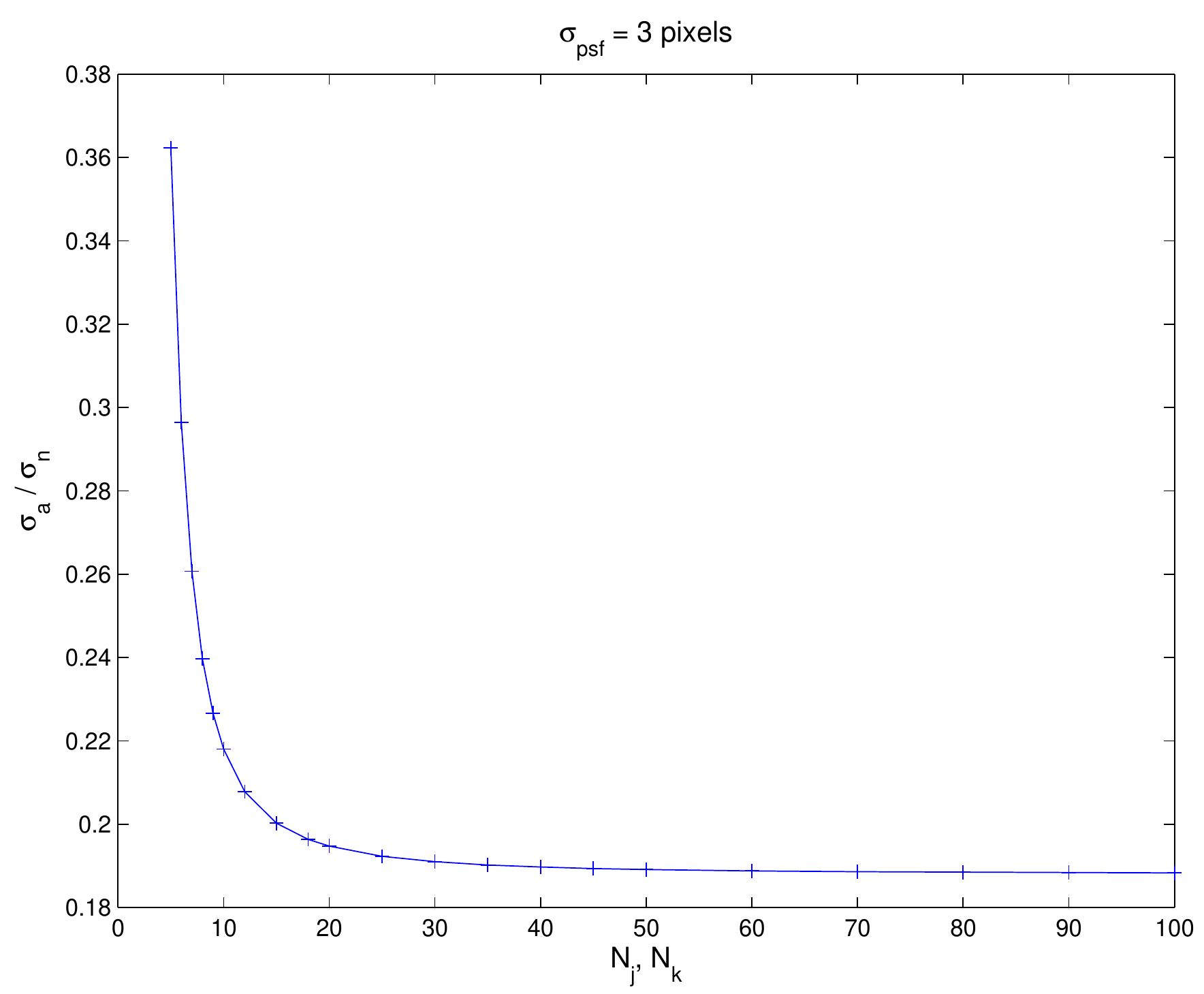}}
        \caption{Standard deviation $\sigma_a$ of the estimated intensity $a$  as provided by MILC in the case of a point source, with a Gaussian profile with dispersion $\sigma_{\rm psf}$ equal to $3$ pixels, as a function of the sizes $N_j = N_k$ of the searching patch. Here, a single map is considered with a background given by a  two-dimensional, one-degree polynomial. Instrumental noise is Gaussian and white with standard deviation $\sigma_n$, The true value of ``$a$'' 
is $1$ in units of $\sigma_n$.}
        \label{fig:area1}
\end{figure*}

\begin{figure*}
        \resizebox{\hsize}{!}{\includegraphics{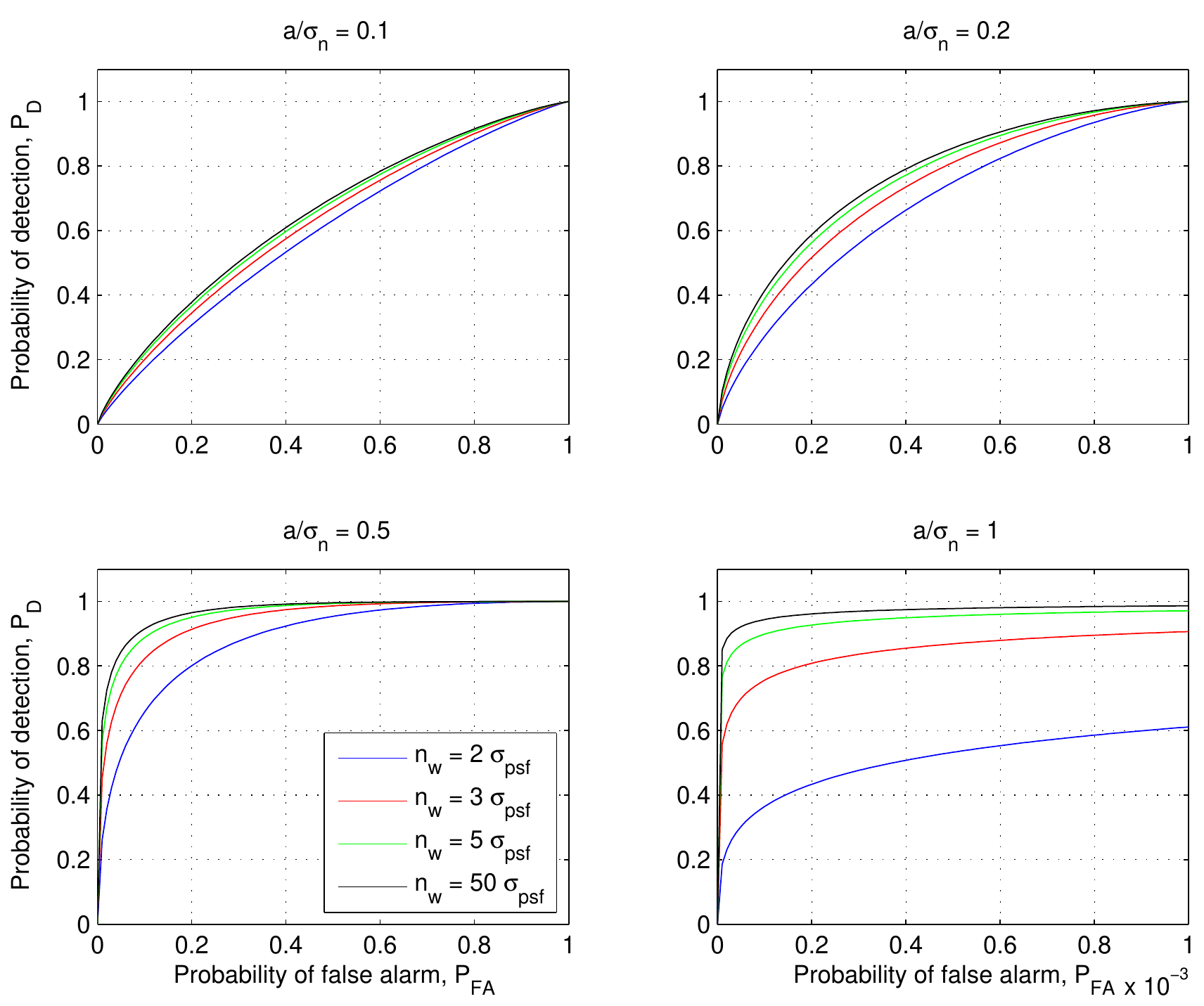}}
        \caption{Relationship between the {\it probability of detection}, $\PD$, vs. the {\it probability of false alarm}, $\PFA$ for the case shown in Fig.~\ref{fig:area1} but for different values of the ratio $a / \sigma_{n}$. Note the different
         scale used for the abscissa in the bottom-right panel.}
        \label{fig:area2}
\end{figure*}

In this section we discuss some practical problems and how they can be addressed. The first is related to the degree $m$ of the polynomial used to approximate the background.
Obviously, the smaller the sky area of interest the lower the degree of the polynomial. For example, considering that the CMB has a coherence scale of about $10~{\rm arcmin}$, it can be reasonably expected that 
with a resolution of $0.1$-$1.0~{\rm arcsec}$ a first-degree polynomial is a good choice. The second question is related to the sizes $N_j$ and $N_k$ of the patch for
testing for the presence of a point source. Two competing requirements arise: on the one hand, $N_j$ and $N_k$ must be as large as possible to reduce errors in estimating the polynomial parameters, on the other, a small size implies that the approximation of the background with a low-degree polynomial is a more reliable operation and, at the same time, that the probability two or more sources being in the same patch $\Xmatb$ is low. 
For illustrative purposes, Fig.~\ref{fig:area1} shows 
the standard deviation $\sigma_a$ of the estimated intensity $a$ as provided by MILC in the case of a point source with a Gaussian profile and a dispersion $\sigma_{\rm psf}$ equal to three pixels. A single map is considered where the 
background is given by a two-dimensional one-degree polynomial, instrumental noise is Gaussian and white with standard deviation $\sigma_n$, and $N_{p_1} $ and $N_{p_2}$ are progressively increased. 
The true value of $a$ is one in unit of $\sigma_n$. The decrease in $\sigma_a$ is evident.  Figure~\ref{fig:area2} shows the relationship between $\PD$ and $\PFA$ for different values of the ratio $a / \sigma_n$. 
These figures clearly show that $N_j$ and $N_k$ lying in the range $3 \sigma_{\rm psf}$-$5 \sigma_{\rm psf}$ is a reasonable compromise.

As shown in the appendix~\ref{sec:shape}, that the exact PSF $\Fmatb$ could not be known or that some of the point sources could be overlapping and/or could have an extended shape,
have no important consequences. However,
another issue arises because the shape of the PFSs changes with observing frequency in practical applications. Widespread practice is to convolve maps with a suited kernel function in order to get a common spatial PSF
$\Fmatb$ for all the frequencies. This operation has the beneficial effect of reducing the standard deviation of the instrumental noise, but at the same time it introduces a spurious spatial correlation in it. Actually, even if neglected, this
is not expected to be critical since  both MILC and MMILC are linear techniques, and the only consequence is a reduction of the efficiency  of the least-squares estimate of the coefficients $\qb$ 
(i.e., the estimate is unbiased but with greater variance). Something similar is also expected for NP-MILC and NP-MMILC that represent the solution of a linear least-squares problem with a quadratic constraint (i.e., both the quantity to minimize and the constraint are smooth functions). In other words, given the above-mentioned reduction in the standard deviation of the noise,
this spurious correlation is not expected to have critical consequences. This is especially true if one takes into account that there are other and more important approximations that make the analysis of data less rigorous (e.g., often the level of instrumental 
noise is only roughly known).

A final question regards whether in practical applications it is more convenient to use the MILC and MMILC algorithms or the NP-MILC and NP-MMILC ones. 
Indeed, MILC and MMILC work optimally only for a specific emission spectrum $\ahb$, a  feature common to other detection techniques such as the {\it matched multifilter} \citep{her12}.
In principle, this is not a critical question. It is sufficient to apply the detection algorithm to a set of prefixed $\ahb$ obtained by grouping sources in broad families - radio flat, radio steep, dusty galaxies of a certain type, etc - and
defining average spectral laws $\ahb$ for each family \citep{ram11, her12}. Such an approach is viable since MILC and  MMILC are fast algorithms, and they require the numerical solution of linear systems containing no more than a few of tens of
linear equations. Moreover, as shown in \citet{ram11},  where a version of MMILC without background subtraction is applied to high-Galactic latitude WMAP maps, strong degradation of the detection capability has to be expected only if the
spectrum of the point sources is quite different from the one for which the MMILC algorithm has been optimized. Of course, this kind of problem can be avoided using NP-MILC and NP-MMILC. However, since they are not optimized for specific
emission characteristics, the price is a lower detection capability for specific spectra. Given the inexpensive computational cost of the four algorithms, the best choice is to try all of them and check the results.

\section{Numerical experiments}

To support the arguments presented above, we present some numerical experiments here with simulated maps at high Galactic latitude (where the Galactic contamination is negligible) that is the region of interest for future 
CMB experiments. Since realistic experimental conditions are not yet available, such simulations are only presented for illustration.

We produced small sky patches of $0.86 ~{\rm deg}^{2}$ at $3''$ angular resolution with several components, 
namely, the CMB and the  Sunyaev-Zel'dovich effects (SZ), both kinetic and thermal. 
To produce these maps we used hydrodynamic/N-body simulations with cosmological parameters that are consistent 
with WMAP parameters for a flat Universe and standard $\Lambda$CDM model, with an equation of state 
for the dark energy component of $w=-1$. The adopted present time density parameters expressed in terms 
of the critical density are $(\Omega_{\rm cdm},\Omega_{\Lambda},\Omega_b)=(0.256, 0.7, 0.044)$, a 
dimensionless Hubble constant of $h=0.71$, and a mean CMB temperature of $T$=2.725 K. 
Adiabatic initial conditions are assumed, a spectral index of $n_s=1$, and full reionization at redshift $7$. 

For the present epoch, we considered a normalization power spectrum of $\sigma_{8}=0.9$ 
and a shape parameter of $\Gamma=0.17$.  
The CMB component is produced with the CAMB code \citep{lew00} 
to obtain the linear CMB power spectrum. The full-sky CMB temperature anisotropy map was generated 
with the HEALPix software \citep{gor05} with ${\rm Nside}=8192$. From this map a small sky 
region was extracted with an area of about 0.86 deg$^{2}$ around the equator, projected on a squared map.
Details about the simulations of the SZ effect components can be found in \citet{das01} and \citet{ram12}. 
The frequencies chosen were $90$, $150$, $250$, $330$, 
$440$, $675$, and $950~{\rm GHz}$, which correspond to the ALMA receiver bands.

All components were co-added, resulting in a final map, $\Delta I_{\rm CMB+SZ}/I$, with a pixel size of $3 ~{\rm arcsec}$.
We use the central part of the maps ($300 \times 300$ pixels) and convolve them for each frequency with a Gaussian PSF with a dispersion of three pixels. To each map a white-noise process has also been added with 
standard deviations $\sigma_{n_i}$ set to $0.12$ time the standard deviation of the values of 
map itself. Finally, $20$ randomly distributed point sources were included with $a_i = 1.7 \sigma_{n_i}$.  In this way, maps with the same {\rm S/N} are obtained.
The values of $\sigma_{n_i}$ and $a_i$  have been arbitrarily chosen to test algorithms under very bad operational conditions, but at the same time to obtain
stable results (i.e. with different realizations of the noise process almost all the sources are correctly detected with no false detections). 

\begin{figure*}
        \resizebox{\hsize}{!}{\includegraphics{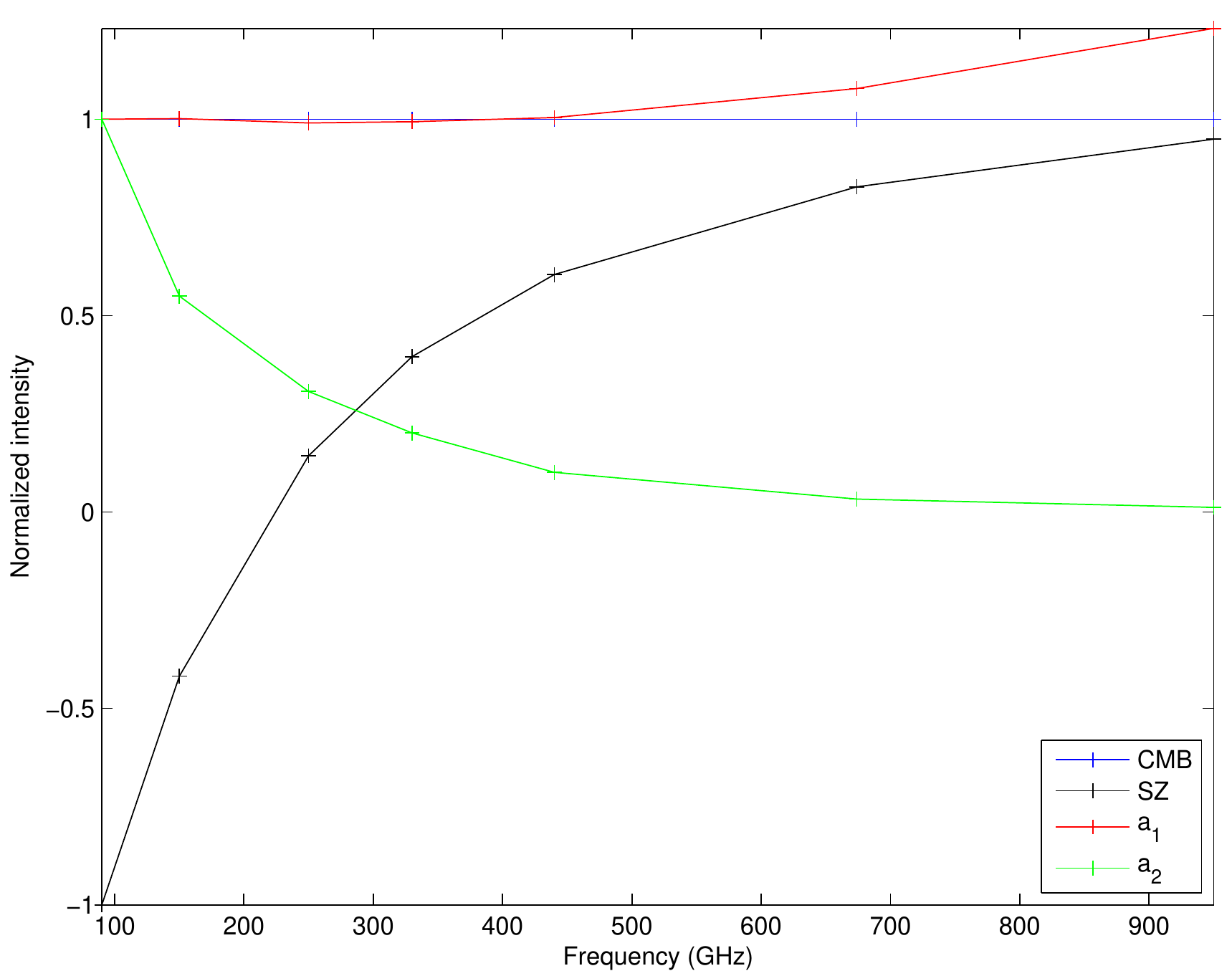}}
        \caption{Comparison of the spectrum of the point sources used in the numerical experiments of Figs.~\ref{fig:detection1b}-\ref{fig:detection2c} with that of CMB and the thermal SZ. All spectra have been normalized in such a way to have 
        absolute intensity equal to $1$ at $90~{\rm GHz}$. Here, $\ab_1$ and $\ab_2$ indicate the point source with spectrum, respectively, similar to and unlike that of CMB used in the numerical experiment (see next figures).} 
        \label{fig:spectra}
\end{figure*}

\begin{figure*}
        \resizebox{\hsize}{!}{\includegraphics{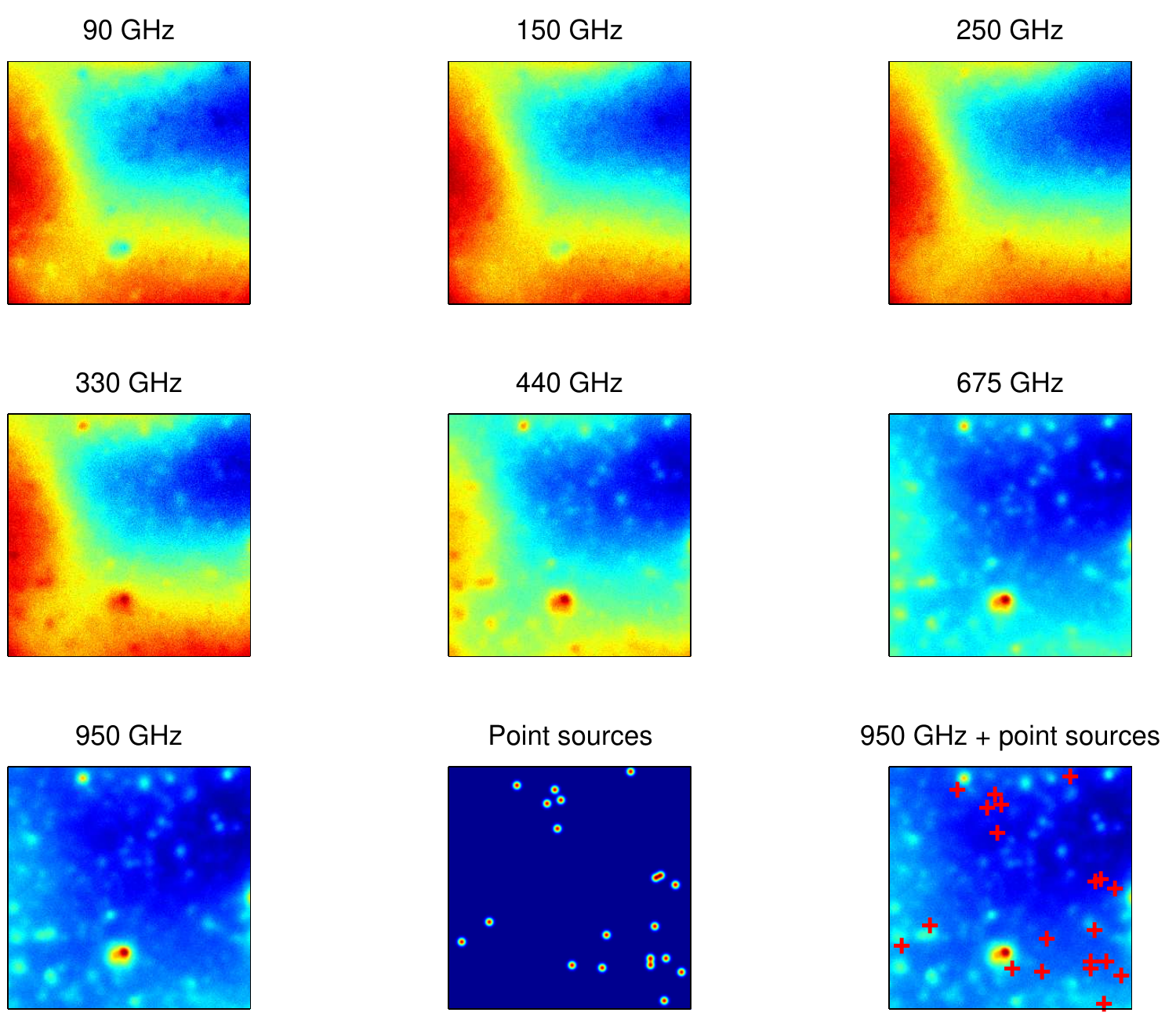}}
        \caption{Simulations of a sky region at high Galactic declination at the ALMA observing frequencies. $20$ randomly distributed point sources with the same intensity have been added. 
	Here, the point sources have a spectrum similar to that of CMB  (see text and curve $\ab_1$ in Fig.~\ref{fig:spectra}). The PSFs are assumed to be Gaussian with a standard 
	deviation of $3$ pixels.  Noise is Gaussian-white with standard deviation set to $0.12$ time the standard deviation of the values in the corresponding  noise-free maps. All of the point sources have the same
         intensity set to $1.7$ times the standard deviation of the noise. The two bottom-right panels show the simulated point sources and their position on the $950~{\rm GHz}$ map. }
        \label{fig:detection1b}
\end{figure*}

\begin{figure*}
        \resizebox{\hsize}{!}{\includegraphics{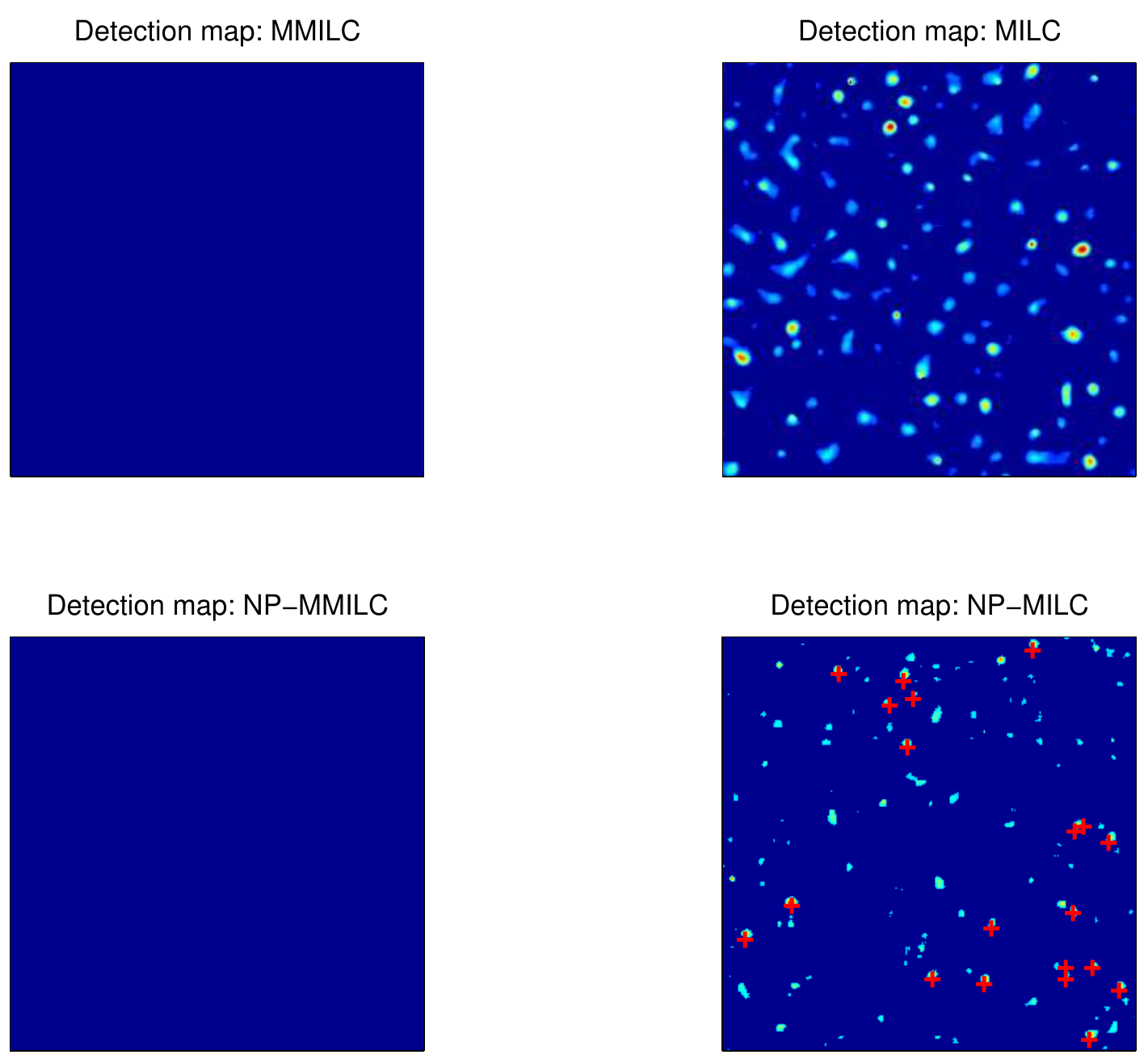}}
        \caption{Results provided by MILC, MMILC, NP-MILC, and NP-MMILC when applied to the maps in Fig.~\ref{fig:detection1b}. The detection threshold has been set to  $4 \sigma_{\rm L}$ for all algorithms except for NP-MMILC
for which a value of $3.5 \sigma_{\rm L}$ has been adopted (see text) and the background has been approximated by a
         two-dimensional polynomial of degree one. The top and bottom left  panels clearly show that both MMILC and NP-MMILC are not able to retrieve point sources in this case. This happens because the spectrum of the point sources has a 
frequency-dependence similar to the CMB and SZ, and therefore the subtraction process gets rid of all of them. The top and bottom right panels show that both MILC and NP-MILC, on the contrary, retrieve all sources and the 
SZ point-like emissions, because it subtracts the underlying diffuse component with the polynomial approximation. For NP-MILC a greater noise contamination of the detection map is evident.
Also notice the detection of two couples of overlapping point sources in the bottom-right and middle-right part of the map.}
        \label{fig:detection1c}
\end{figure*}

The simulated experimental scenario corresponds to an adverse situation of rather low {\rm S/N} and, since $\sigma_{n_i}$ increases with frequency, with a spectrum $\ahb$ (see curve $\ab_1$ in Fig.~\ref{fig:spectra}) 
that mimics that of the CMB plus SZ background (i.e. $\ahb$ is close to the {\it nullspace} of $\Mb$, or $\Mb \ahb \approx \zerob$).
Figure~\ref{fig:detection1b} displays the simulated maps.  We note that the point sources are not even visible, and they are by far exceeded by the SZ point-like emission.
Figure~\ref{fig:detection1c} shows the results obtained
with the four algorithms presented above.  For MILC, MMILC, and NP-MILC the detection threshold has been set to
$4 \sigma_{\rm L}$ whereas a value of $3.5 \sigma_{\rm L}$ has been used  for NP-MMILC. Background has been approximated by a two-dimensional first-degree polynomial. A sliding square window of $19 \times 19$ pixels
has been adopted for the local search of point sources. As expected, the MMILC and NP-MMILC do not work. On the other hand, MILC and NP-MILC
have effectively removed the CMB, as well as the SZ diffuse components, and correctly detected all the point sources in the map (see Fig.~\ref{fig:detection2d}). However, many of the SZ point-like emissions are also present. 

\begin{figure*}
        \resizebox{\hsize}{!}{\includegraphics{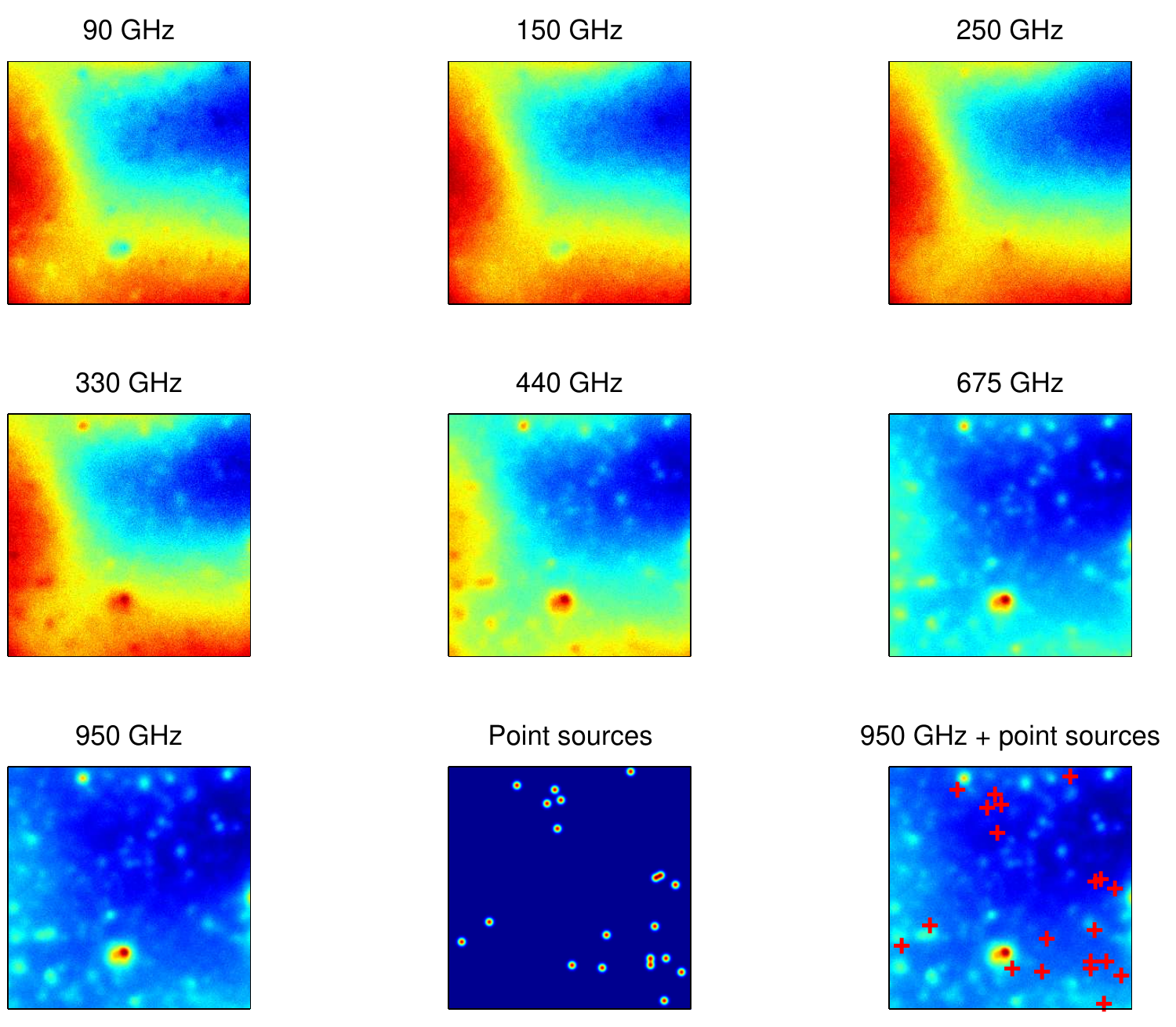}}
        \caption{Simulations of a sky region at high Galactic declination at the ALMA observing frequencies, with $20$ randomly distributed point sources with the same intensity added. In this case the point sources have a spectrum given by  curve $\ab_2$ in Fig.~\ref{fig:spectra}.
        The PSFs are assumed to be Gaussian with a standard deviation of $3$ pixels. Noise is Gaussian-white with standard deviation set to $0.12$ times the standard deviation of the values in the corresponding  noise-free maps.
The PSFs are assumed to be Gaussian with a standard deviation of $3$ pixels. The two bottom-right panels show the simulated point sources and their position on the $950~{\rm GHz}$ map.}
        \label{fig:detection2b}
\end{figure*}

\begin{figure*}
        \resizebox{\hsize}{!}{\includegraphics{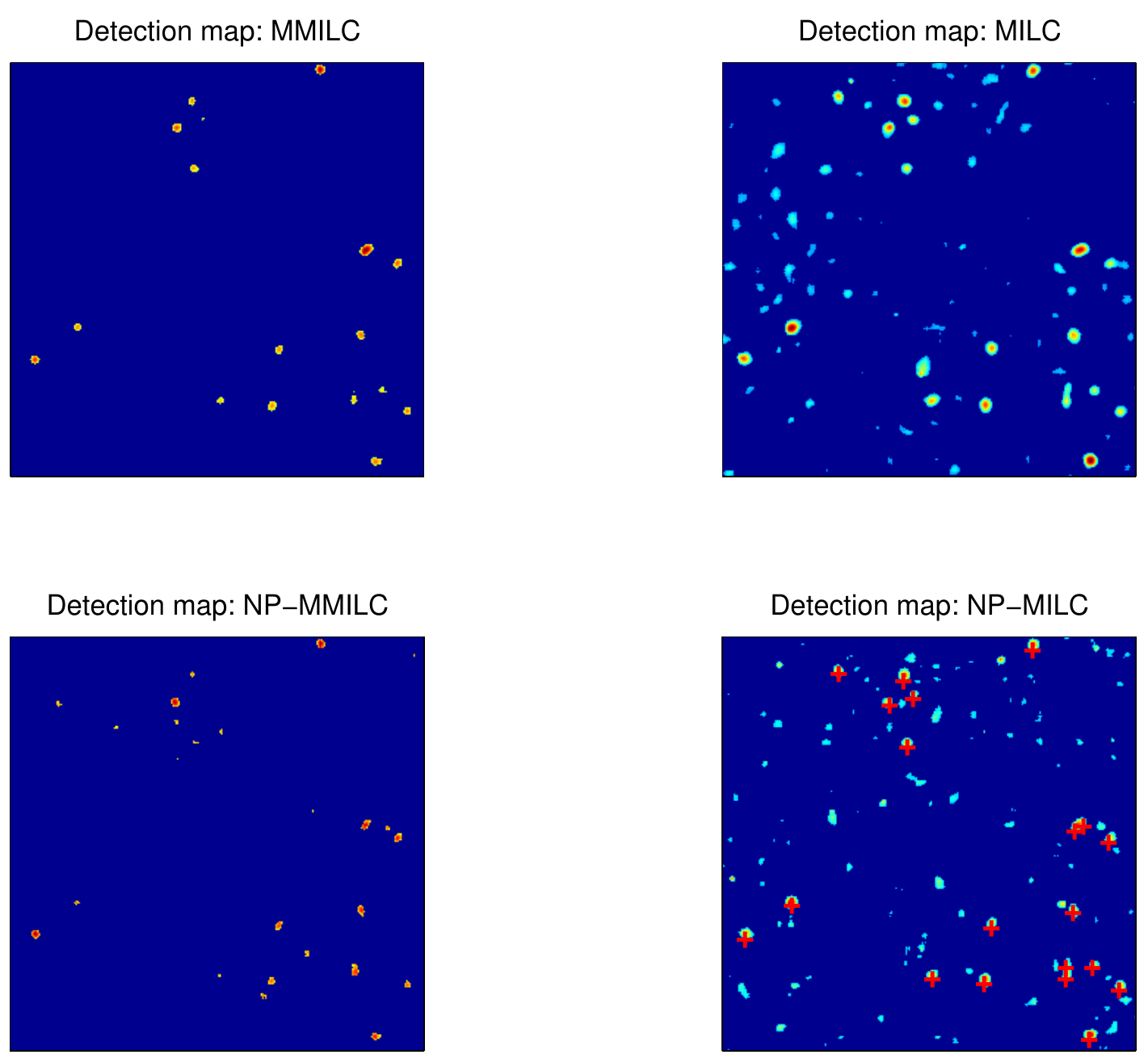}}
        \caption{Results provided by MILC, MMILC, NP-MILC and NP-MMILC when applied to the maps in Fig.~\ref{fig:detection2b}. The detection threshold has been set to  $4 \sigma_{\rm L}$ for all algorithms except for NP-MMILC
for which a value of $3.5 \sigma_{\rm L}$ has been adopted (see text), and the background has been approximated by a two-dimensional polynomial of degree one. The top and bottom left panels clearly show that in this case both 
MMILC and NP-MMILC miss only one point source and get rid of the SZ point-like emissions. A greater noise contamination in the detection map of NP-MMILC is also evident. The top and bottom  right panels show
        that MILC and NP-MILC retrieve all point sources but also the SZ point-like emissions. This is a consequence of subtracting the underlying diffuse component with the polynomial approximation. Also for NP-MILC a greater noise
contamination of the detection map is evident, and notice the detection of two couples of overlapping point sources in the bottom-right and middle-right part of the map.}
        \label{fig:detection2c}
\end{figure*}

\begin{figure*}
        \resizebox{\hsize}{!}{\includegraphics{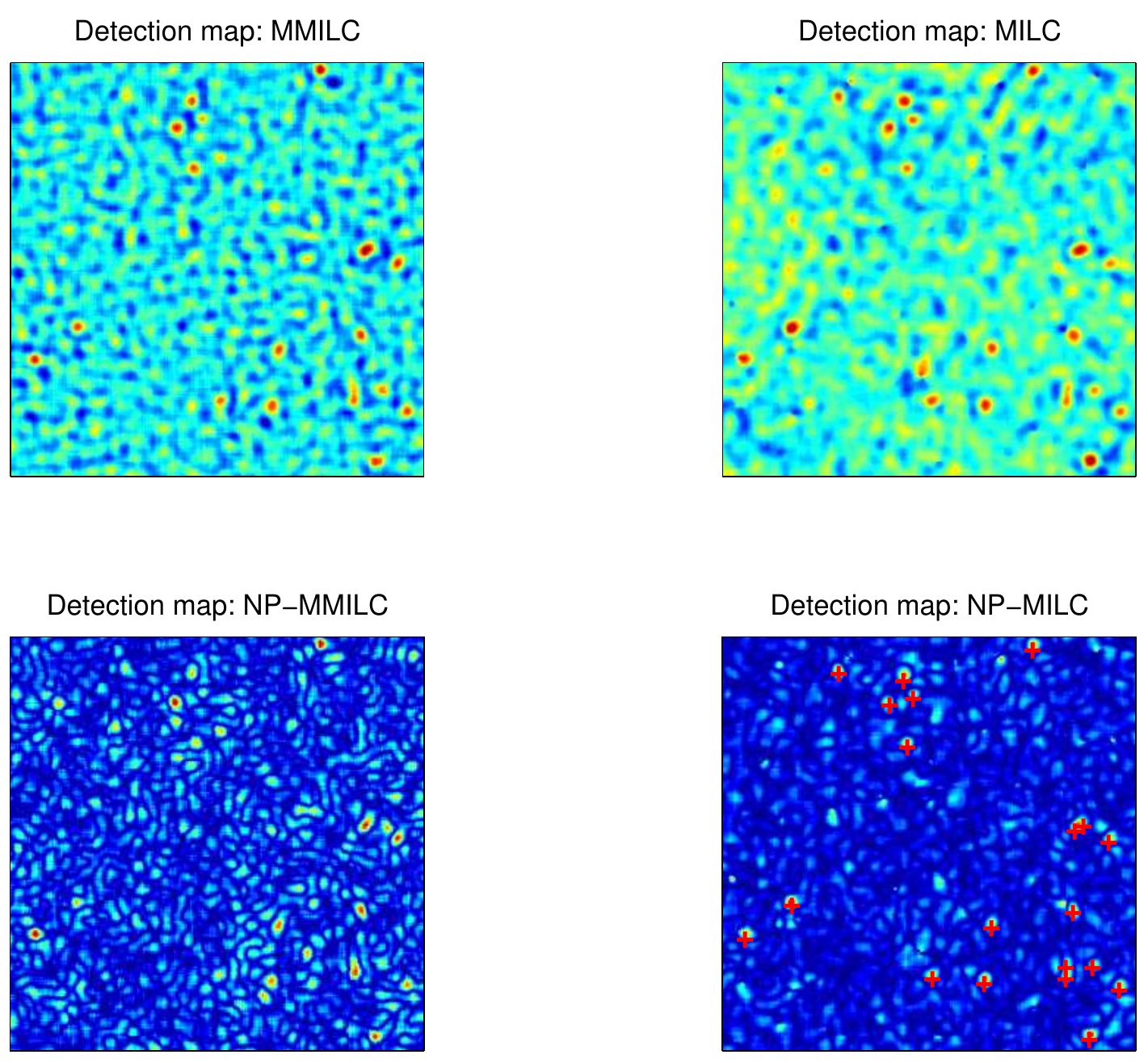}}
        \caption{As in Fig.~\ref{fig:detection2c} with the difference that maps has not been thresholded. This is only to show that the CMB and the extended SZ components have been effectively removed by the polynomial approximation of the background.}
        \label{fig:detection2d}
\end{figure*}

The situation greatly improves if $\{ a_i \}= [2.00,  1.00, 0.60, 0.40, 0.20, 0.06, 0.02]^T \odot \sigmab_n$ which simulates an emission spectrum decreasing with frequency
(see curve $\ab_2$ in Fig.~\ref{fig:spectra}). In this way $\Mb \ahb \ne \zerob$. Figure~\ref{fig:detection2b} shows the maps
corresponding to this case. As is visible in Fig.~\ref{fig:detection2c}, now both MMILC and NP-MMILC are able to correctly detect all the point sources except one, as well to completely remove the CMB and SZ contamination. 
We note the detection of two couples of overlapping point sources in the bottom right hand and middle right hand parts of the map.
However, as expected, the NP-MMILC map presents various false detections due to noise (the detection threshold has been set to  $3.5 \sigma_{\rm L}$ in order to obtain the same number of true detections as MMILC).
A point to stress is that, in the present high Galactic latitude experiment, the diffuse components $\{ \gb_i \}$ can be set to zero. As a result, in models~(\ref{eq:problem1}) and (\ref{eq:problem1a}) it should have been possible
to set $\Lb = \fb$. (i.e., to avoid the polynomial approximation of the background). However, the use of the  general model permits the stability of MMILC and NP-MMILC to be tested for high level of noise.
 
\section{Summary and conclusions}

In this paper four algorithms,  the {\it modified} ILC (MILC), the {\it modified multiple} ILC (MMILC), the {\it nonparametric}-MILC (NP-MILC), and the {\it nonparametric}-MMILC (NP-MMILC), have been presented for detection
of extragalactic point sources in multifrequency, very high-resolution CMB maps. In particular, MMILC and NP-MMILC
make use of the {\it a-priori} information about the spectral properties of the CMB and SZ with MMILC tailored to detecting extragalactic point sources with a given spectrum that has to be different from that of these emissions. 
The other two algorithms are suited to detecting the extragalactic point sources with similar spectra to the CMB or SZ with MILC tailored to the specific spectra.
The main property of the proposed algorithms is that they do not require any {\it a-priori} knowledge
of the statistical characteristics of spatial distribution of the diffuse emissions that contribute to the microwave background. Indeed, these can be locally approximated with a low-degree, two-dimensional polynomial.
The two proposed sets of algorithms used in conjunction are effective in detecting extragalactic point sources independently of the spectral characteristics of their emission.
Their potential performance has been illustrated with some numerical experiments.

\begin{acknowledgements} 
E. P. Ramos is supported by grant POPH-QREN-SFRH/BD/45613/2008, from FCT (Portugal). 
E. P. Ramos and R. Vio thank ESO for its hospitality and support through the DGDF funding program.
\end{acknowledgements}

\clearpage

\clearpage
\appendix

\section{Some additional questions}   \label{sec:shape}

The detection techniques presented in Sec.~\ref{sec:detection} are based on the assumptions that a) the true PSF $\Fmatb$ is known; b) all the sources have a point-like shape (i.e. the shape of the PSF); c) all the sources are isolated.
In practical applications, however, the PSF has to be estimated, some sources can have an extended shape, and overlapping is possible. Here, we show that all this can be expected to have only secondary
consequences. In this respect, for sake of ease in the formalism, we start by considering an experimental one-dimensional signal $\xb = a \fb + \nb$. Here, $\fb$ is the one-dimensional PSF and $\nb$ an additional white noise. 
In practice, $\xb$ is due to a point source with amplitude $a$ embedded in noise. As is well known, the least-squares fit estimate of $a$ provides the solution 
\begin{equation}
\widehat{a}= \frac{\fb^T \xb}{\fb^T \fb}. \label{eq:esta}
\end{equation}
At this point, we point out that in detection problems the test statistic $T(\xb)  = \fb^T \xb$  \citep[e.g. see ][]{kay98} is used where $\fb$ is the well known {\it matched filter} (MF). In other words, unless of a constant factor, 
the estimate $\widehat{a}$ is identical to the test statistic $T(\xb)$.
Indeed, the rhs of Eq.~(\ref{eq:esta}) is nothing else that the normalized correlation of $\xb$ with $\fb$. This means that in the present context, the least squares fit and the matched filter are two equivalent techniques.
This allows us to analyze the characteristic of the least-squares fit using the body of the theory of the linear filters. 
Therefore,  $\fb$ can be interpreted as a linear, typically  low-pass, discrete filter with $N_p$ entries that is made to slide across an array $\xb \equiv \{ x_i \}_{i=1}^{N_{\xb}}$ with $N_{\xb} \gg N_p$. In this way, for each
element $x_i$ one can obtain the quantity  $\widehat{a}_i=T(\xb_i) / (\fb^T \fb)$, where $\xb_i \equiv \{ x_j \}_{j=i}^{i+N_p-1} $. By means of the {\it discrete Fourier transform} (DFT) it is possible to compute $\ahb \equiv \{ \widehat{a}_i \}_{i=1}^{n_{\xb}}$ via
\begin{equation}
\ahb= \frac{{\rm IDFT}[ \tilde{\xb} \odot \tilde{\fb}^*]}{\fb^T \fb},
\end{equation}
with $ \tilde{x}$ and $\tilde{f}$ the DFT of $x$ and the zero padded version of $f$, ${\rm IDFT}[.]$ the inverse DFT operator and  symbols ``$~{}^*~$'', ``$\odot$'' indicating the {\it complex conjugate} and the {\it point-wise multiplication}, respectively.
This implies that the statistical characteristics of $\ahb$ can be analyzed by means of the spectral properties of $\fb$. These considerations can be trivially extended to the two-dimensional case.

\begin{figure*}
        \resizebox{\hsize}{!}{\includegraphics{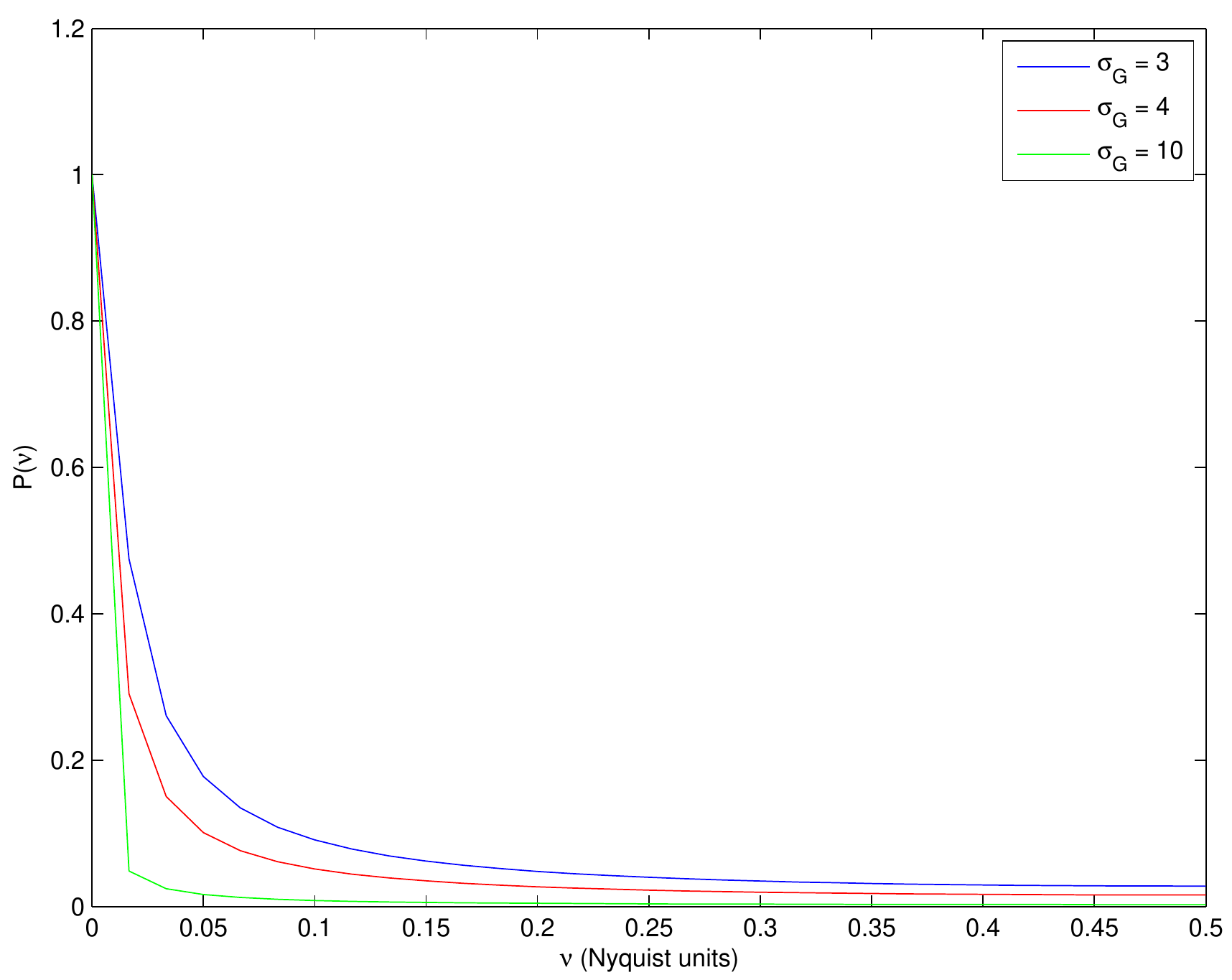}}
        \caption{Central slice of the spectrum $P(\nu)$ of three bivariate, circularly symmetric, Gaussian  PSFs with dispersion $\sigma_G = 3, 4, 10$, respectively. Frequency $\nu$ is in {\it Nyquist units}. }
        \label{fig:psf}
\end{figure*}

\begin{figure*}
        \resizebox{\hsize}{!}{\includegraphics{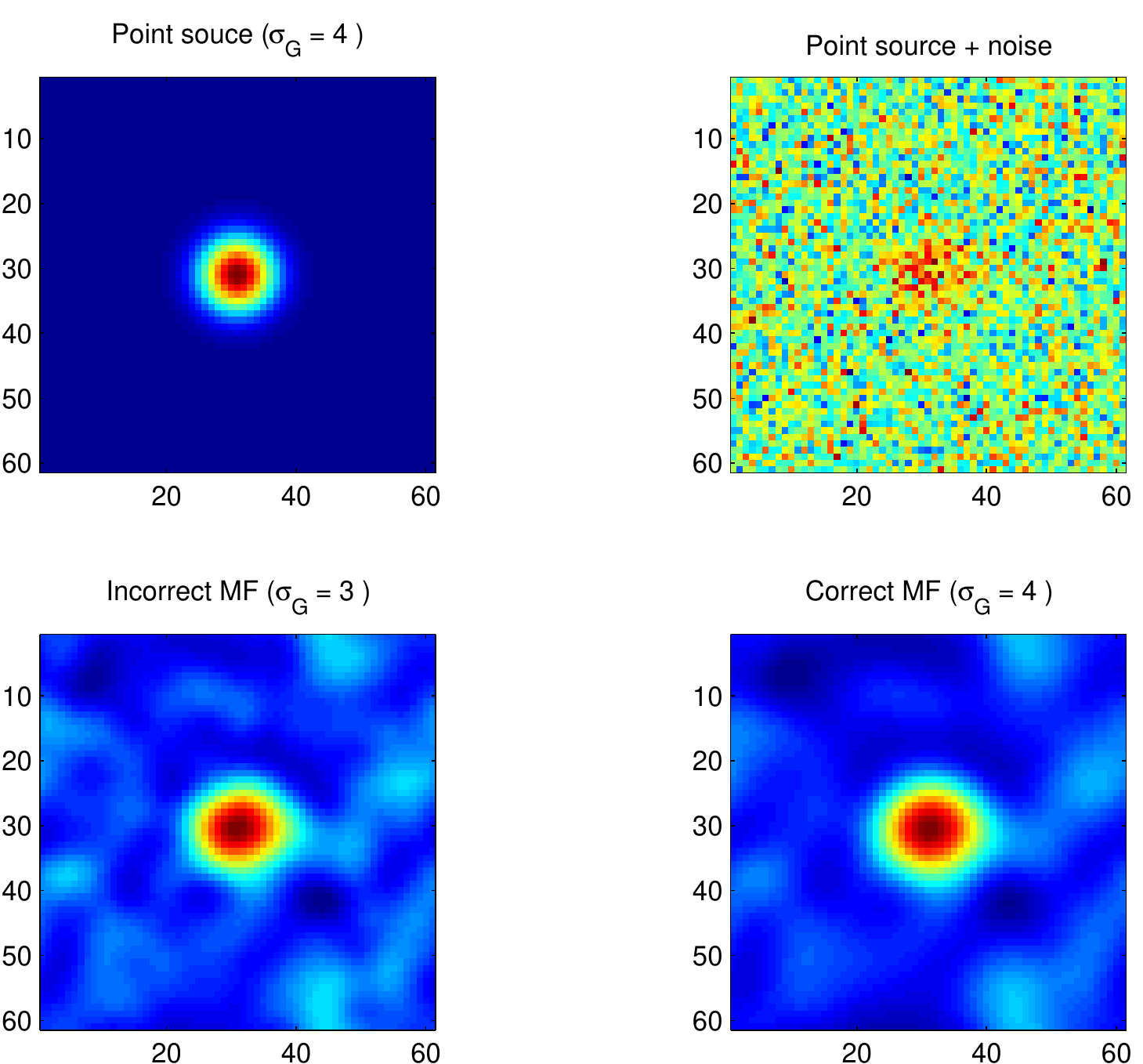}}
        \caption{Upper left panel: image of the original point source when the PSF is a bivariate, circularly symmetric, Gaussian PSF with dispersion $\sigma_G = 4$ pixels. Upper right panel: original point source added  with a white-noise with 
standard deviation equal to half the peak value of the source itself; noisy image filtered with an improper matched filter (MF)  which has a Gaussian shape and $\sigma_G = 3$ pixels; bottom right panel: noisy image filtered with the correct MF.}
        \label{fig:gauss1}
\end{figure*}

\begin{figure*}
        \resizebox{\hsize}{!}{\includegraphics{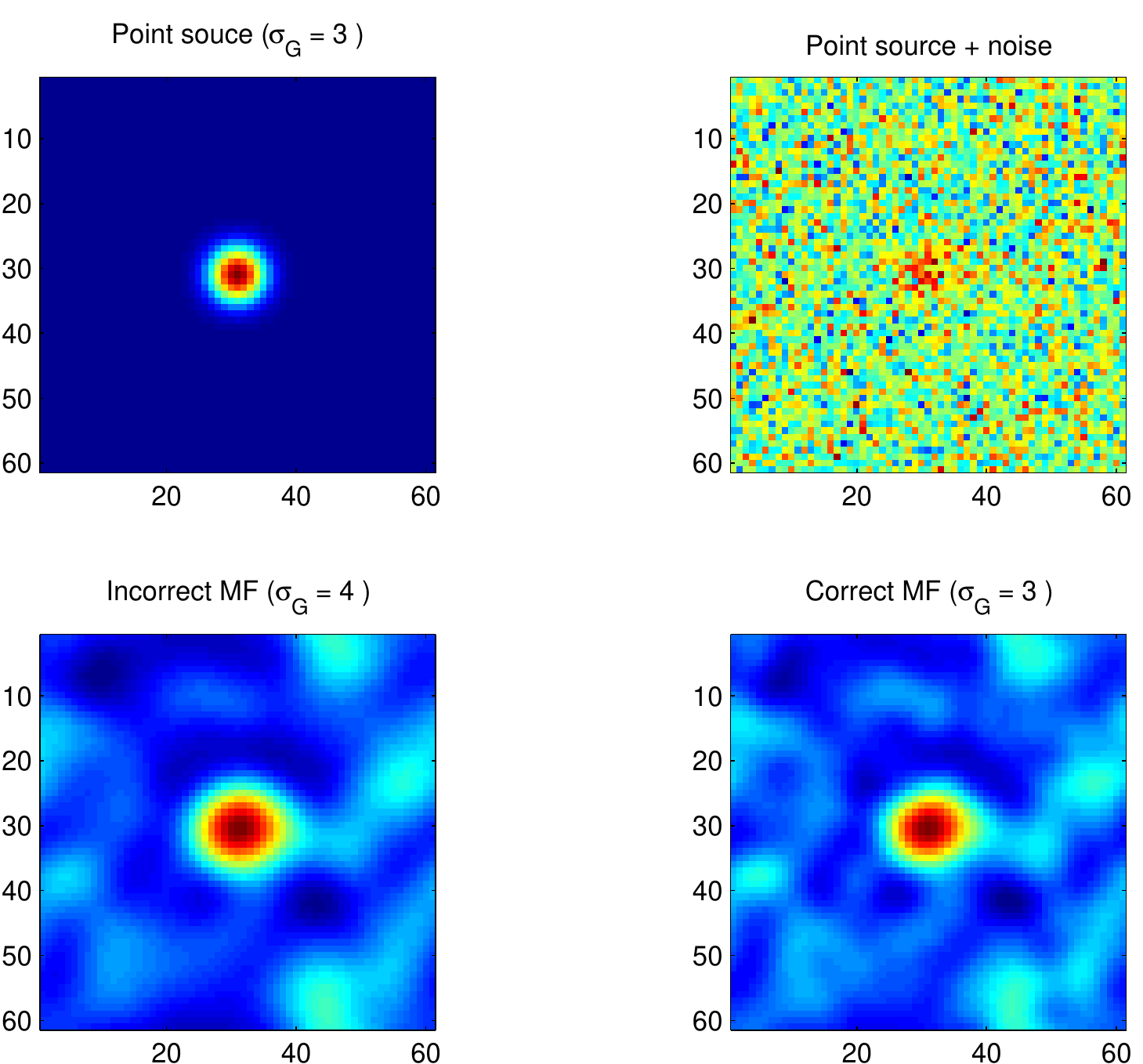}}
        \caption{Upper left panel: image of the original point source when the PSF is a bivariate, circularly symmetric, Gaussian PSF with dispersion $\sigma_G = 3$ pixels. Upper right panel: original point source added  with a white noise with 
standard deviation equal to half the peak value of the source itself. Bottom left panel:  noisy image filtered with an improper matched filter (MF) that has a Gaussian shape and $\sigma_G = 4$ pixels. Bottom right panel: 
noisy image filtered with the correct MF.} 
        \label{fig:gauss2}
\end{figure*}

Here, it is useful to present some simple examples to see that the three facts above are actually not important. Concerning point a), Fig.~\ref{fig:psf} shows a central slice  $P(\nu)= |  \tilde{\fb} |$ of the spectrum
of three two-dimensional, circularly symmetric, Gaussian PSFs with dispersion
$\sigma_G = 3$ and $4$ pixels, respectively. It is clearly visible that the value of $\sigma_G$ determines the bandwidth of $ \tilde{\fb}$ in the sense that the larger $\sigma_G$, the stronger the filtering action. This means that,
if a point source with $\sigma_G = 4$ pixels is filtered assuming a PSF with $\sigma_G=3$  (i.e. with an error of $25\%$), the only consequence is a slighter reduction of the noise with respect to what is obtainable with the correct PSF. In contrast,
if the point source has $\sigma_G=3$ and a PSF with $\sigma_G=4$ is assumed (i.e. with an error of $33\%$), a stronger reduction of the noise is obtained. In this case, however, part of the signal of interest is also filtered out, again with a  
slighter reduction of the detection capability with respect to what is obtainable with the correct PSF. These two examples are shown in Figs.~\ref{fig:gauss1}-\ref{fig:gauss2}.

\begin{figure*}
        \resizebox{\hsize}{!}{\includegraphics{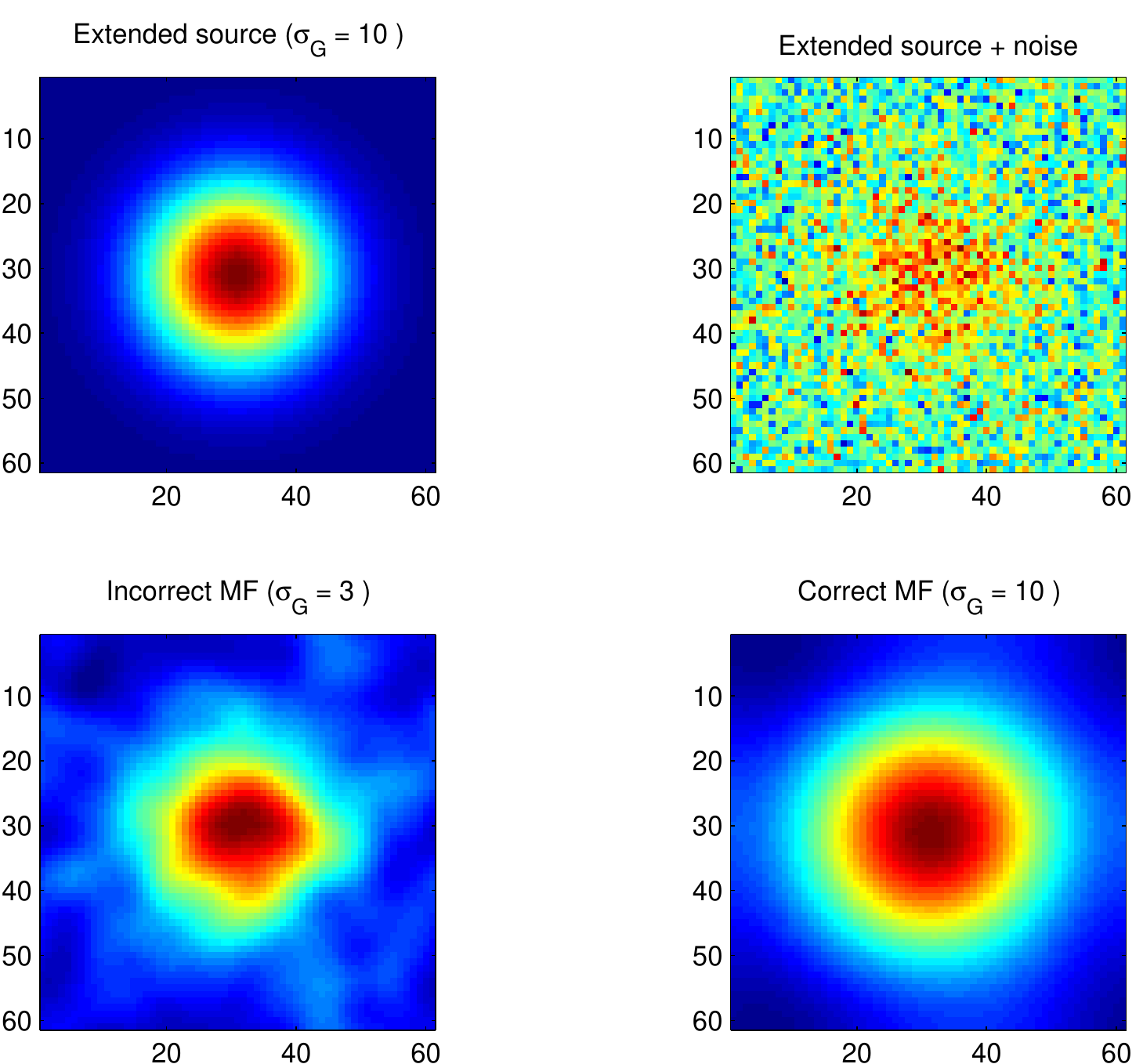}}
        \caption{Upper left panel: image of an extended point source with a circularly symmetric bivariate Gaussian shape with dispersion $\sigma_g = 10$ pixels.  Upper right panel: original point source added  with a white noise with 
standard deviation equal to half the peak value of the source itself. Bottom left panel: noisy image filtered with an improper matched filter (MF), which has a Gaussian shape and $\sigma_G = 3$ pixels. Bottom right panel: noisy image filtered with the correct MF.} 
        \label{fig:gauss3}
\end{figure*}

Similar arguments hold for the point b) when the source has an extended shape. In this case, however, since the PSF is much ``{\it narrower}'' than the source, the noise filtering will be much less effective. However, as shown by the example
in Fig.~\ref{fig:gauss3}, where an extended object with Gaussian shape, $\sigma_G = 10$, and a Gaussian PSF with $\sigma_G=3$ pixels are considered, this does not mean that detection is not possible, but rather that it is less effective.

\begin{figure*}
        \resizebox{\hsize}{!}{\includegraphics{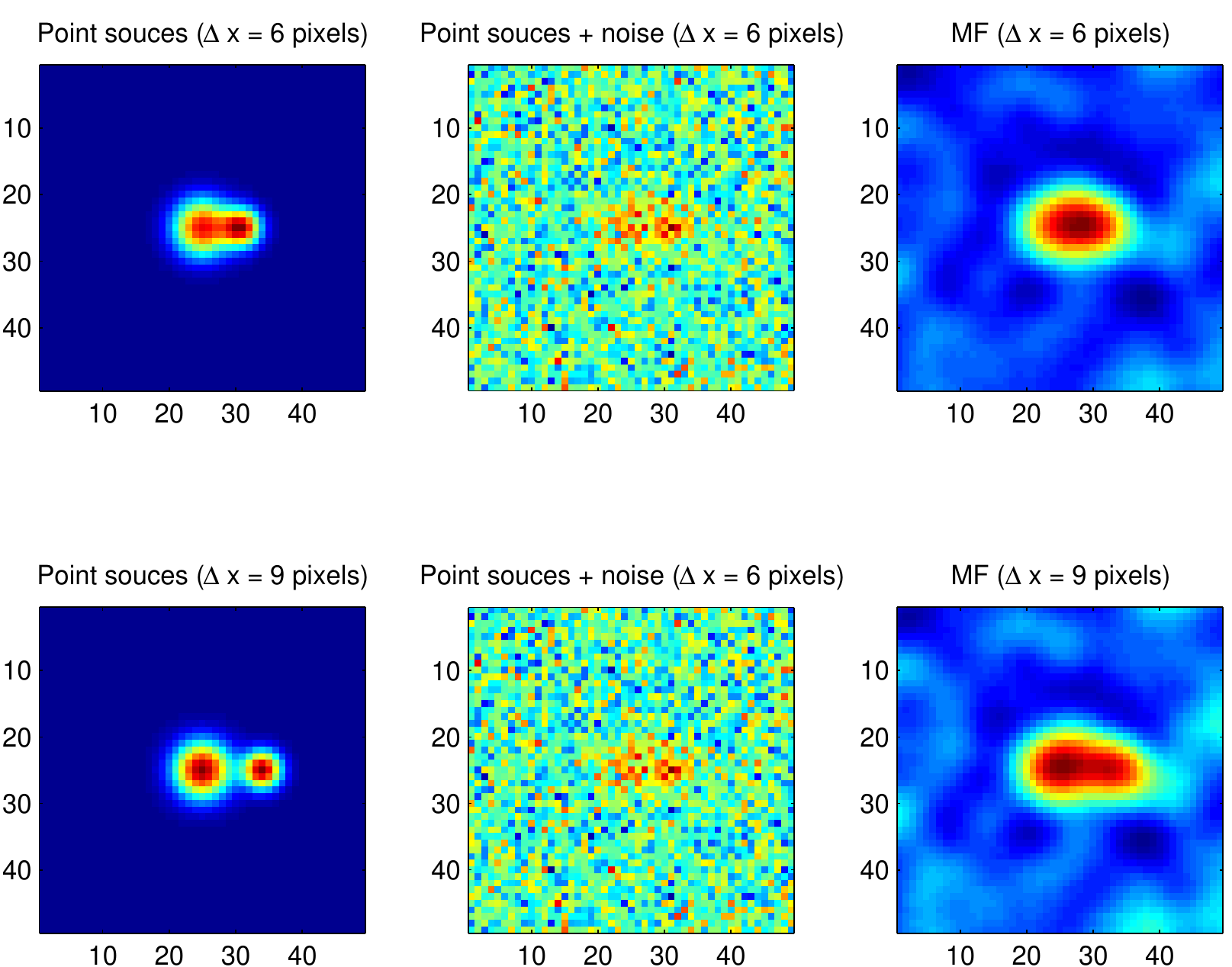}}
        \caption{Results obtainable with the matched filter (MF) in the case of two identical overlapping point sources with shape given by a bivariate, circularly symmetric, Gaussian PSF  with $\sigma_G = 3$ pixels, 
when their peaks are $6$ and $9$ pixels apart. A white noise is added with standard deviation equal to half the peak value of the sources.} 
        \label{fig:gauss4a}
\end{figure*}

Finally, concerning point c), i.e. the case of two close point sources, the situation does not change very much. The only consequence is that the filtered point sources will
appear more overlapped than the unfiltered ones (see Fig.~\ref{fig:gauss4a}).

\begin{figure*}
        \resizebox{\hsize}{!}{\includegraphics{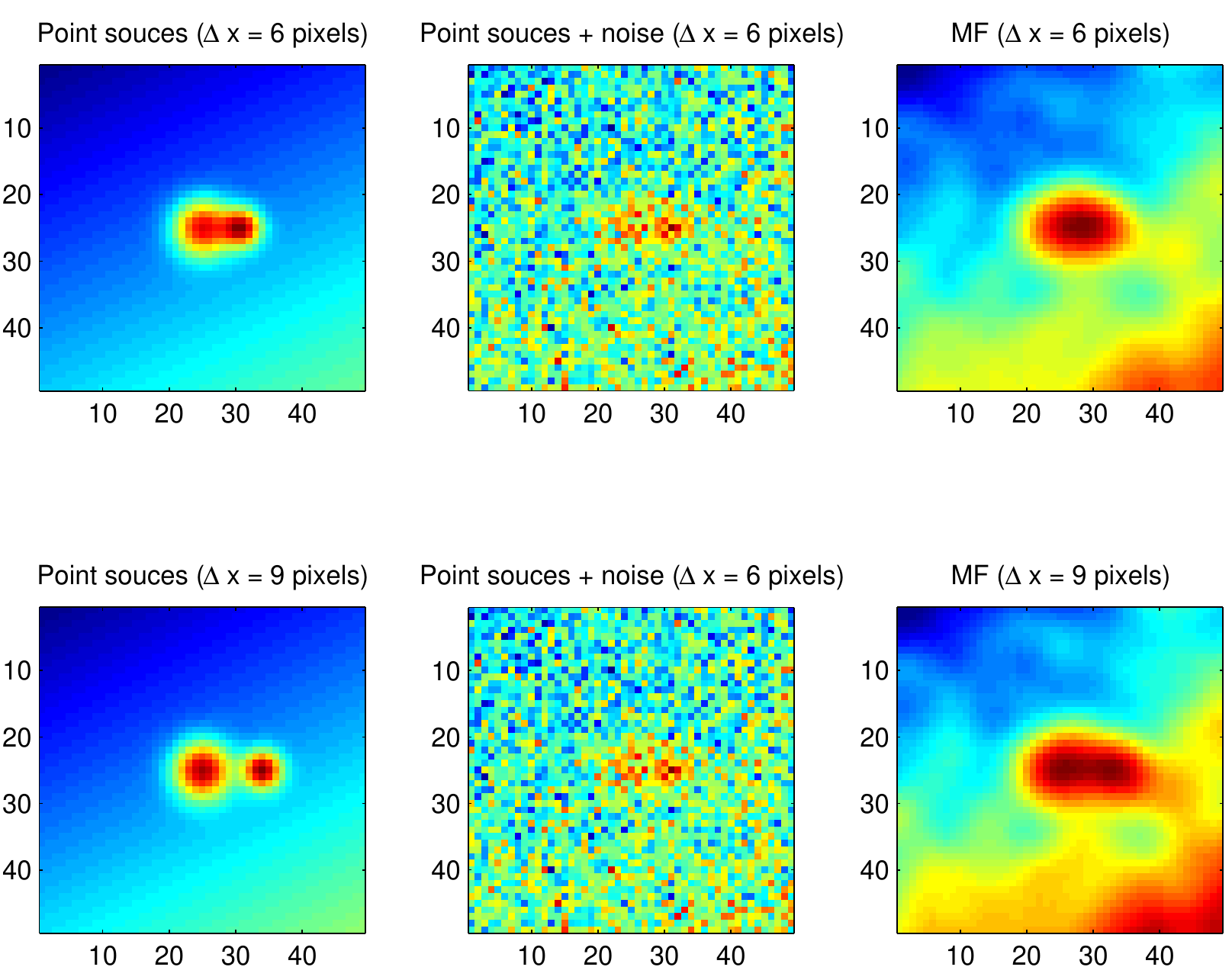}}
        \caption{Results obtainable with the matched filter (MF) in the case of two identical overlapping point sources with shape given by a bivariate, circularly symmetric, Gaussian PSF  with $\sigma_G = 3$ pixels, 
when their peaks are $6$ and $9$ pixels apart. A first-degree, two-dimensional polynomial background and a white noise with standard deviation equal to half the peak value of the sources are added.} 
        \label{fig:gauss4b}
\end{figure*}

In the cases examined above, the presence of a smooth background $\pb$  has not been considered. However, again, things can be expected not to change very much. Indeed,  Eq.~(\ref{eq:esta}) becomes 
\begin{equation}
\widehat{a}= \frac{\fb^T (\xb - \pb) }{\fb^T \fb}. \label{eq:estaa}
\end{equation}
Now, if instead of $\pb$ the result of a least-squares polynomial fit  $\phb$ is used, then one obtains
\begin{equation}
\widehat{a}= \frac{\fb^T \xb }{\fb^T \fb} + \frac{\fb^T \dhb }{\fb^T \fb}. \label{eq:estaaw}
\end{equation}
with $\dhb=\phb-\pb$ a smooth function. In other words, a map is produced where a smoothed point source is superimposed to a smoothed (presumably weak) background and noise is reduced. 
This is visible in Fig~\ref{fig:gauss4b}, which shows the results obtained for a situation similar to that of Fig.~\ref{fig:gauss4a} when a first-degree, two-dimensional polynomial background is present. 

To understand why it is reasonable to assume that these facts are not critical for MMILC, it is useful to interpret this method as done in Sec.~\ref{sec:mmilc}; i.e,
a sequential least-squares fit of a point spread function (PSF) overlapped to a two-dimensional polynomial background on the original data for each frequency, followed by a constrained ILC 
on the residuals. Since, each of the fits is not very sensitive to the issues mentioned above, it is reasonable to assume that the same is valid for their linear combination. Something similar holds for MILC, NP-MILC, and
NP-MMILC. These arguments can also be extended with very good approximation to the case that, as in Sec.~\ref{sec:detection}, the background is computed for each position of the sliding detection window. Indeed, since the window is made to slide  
one pixel at a time, for close pixels the least-squares fit is done using essentially the same data. The experiments concerning MILC and NP-MILC in Figs.~\ref{fig:detection1b}-\ref{fig:detection2c} confirm this.


\end{document}